\documentclass[%
 reprint,
 superscriptaddress,
preprintnumbers,
nofootinbib,
 amsmath,amssymb,
 aps,
]{revtex4-1}

\usepackage{amsmath}
\usepackage{amssymb}
\usepackage{amsthm}
\usepackage{MnSymbol}
\usepackage{nicefrac}
\usepackage{amsfonts}
\usepackage{dsfont}

\usepackage[markup=nocolor]{changes}
\usepackage{natbib}

\usepackage{graphicx}
\usepackage{epstopdf}

\usepackage[normalem]{ulem}

\usepackage{dcolumn}
\usepackage{bm}

\usepackage{color}
\usepackage[usenames,dvipsnames,svgnames,table]{xcolor}
\usepackage{slashed}
\definecolor{mygreen}{rgb}{0, 0.6, 0}
\usepackage{empheq}

\newcommand{\bea}{\begin{eqnarray}}
\newcommand{\eea}{\end{eqnarray}}
\newcommand{\be}{\begin{equation}}
\newcommand{\ee}{\end{equation}}

\def\s0#1#2{\mbox{\small{$ \0{#1}{#2} $}}}
\def\0#1#2{\frac{#1}{#2}}

\begin{document}

\title{Viability of quantum-gravity induced ultraviolet completions for matter}
 
 \author{Astrid Eichhorn}
   \email{a.eichhorn@thphys.uni-heidelberg.de}
\affiliation{Institut f\"ur Theoretische
  Physik, Universit\"at Heidelberg, Philosophenweg 16, 69120
  Heidelberg, Germany}
\author{Aaron Held}
\email{a.held@thphys.uni-heidelberg.de}
\affiliation{Institut f\"ur Theoretische
  Physik, Universit\"at Heidelberg, Philosophenweg 16, 69120
  Heidelberg, Germany}

\begin{abstract}
We highlight how the existence of an ultraviolet completion for interacting Standard-Model type matter puts constraints on the viable microscopic dynamics of asymptotically safe quantum gravity within truncated Renormalization Group flows. A first constraint -- the weak-gravity bound -- is rooted in the destruction of quantum scale-invariance in the matter system by strong quantum-gravity fluctuations. A second constraint arises by linking Planck-scale dynamics to the dynamics at the electroweak scale. Specifically, we delineate how to extract a prediction of the top quark mass from asymptotically safe gravity and stress that a finite top mass could be difficult to accommodate in a significant part of the gravitational coupling space.
\end{abstract}

\pacs{Valid PACS appear here}

\maketitle

\section{Connecting quantum gravity to the electroweak scale}
Observational guidance on the route to quantum gravity is notoriously elusive. For most ``smoking-gun" signals of quantum gravity, the Planck scale would have to be accessible by controlled experiments, as effects are typically suppressed by the energy scale of the experiment over the Planck scale. Here, we discuss how to restrict the microscopic quantum-gravity dynamics by bridging the gap in the hierarchy of scales between Planck and electroweak scale in terms of a Wilsonian Renormalization Group (RG) flow. Recovering experimentally tested particle physics at the electroweak scale restricts the form of the microscopic dynamics including quantum gravity. In particular, the microscopic model could even be a fundamental quantum field theory of gravity and matter within the asymptotic-safety paradigm \cite{Weinberg:1980gg}. That scenario generalizes the success-story of asymptotic freedom -- where a free RG fixed point underlies quantum scale-invariance in the ultraviolet (UV) -- to an \emph{interacting} UV complete theory. Based on the groundbreaking work of Reuter \cite{Reuter:1996cp}, compelling hints for asymptotic safety in pure gravity have been found
\cite{Reuter:2001ag,Lauscher:2001ya,Lauscher:2002sq,Litim:2003vp,Fischer:2006fz,
Codello:2006in,Machado:2007ea,Codello:2008vh,Eichhorn:2009ah,Manrique:2009uh,
Benedetti:2009rx,Eichhorn:2010tb,Groh:2010ta,Manrique:2010mq,Manrique:2010am,
Manrique:2011jc,Benedetti:2012dx,Christiansen:2012rx,Dietz:2012ic,Falls:2013bv,
Christiansen:2014raa,Becker:2014qya,Falls:2014tra,Eichhorn:2015bna,
Demmel:2015oqa,Christiansen:2015rva,Gies:2015tca,Morris:2016spn,Percacci:2016arh,Gies:2016con,Ohta:2016npm,Henz:2016aoh,Biemans:2016rvp,Pagani:2016dof,Christiansen:2016sjn,Denz:2016qks,Falls:2017cze,Gonzalez-Martin:2017gza,Houthoff:2017oam}, see, e.g., \cite{ASreviews} for reviews and \cite{cosmology,Falls:2010he} for possible consequences in astrophysics and cosmology. For asymptotically safe models without gravity see, e.g., \cite{Gies:2003ic,Gies:2009hq,Braun:2010tt,AS4d}.

By including matter, we derive strong hints for two structurally different constraints on the gravitational coupling space: In the UV, we find a weak-gravity bound: A quantum-gravity induced fixed point for matter can only exist for sufficiently weak quantum-gravity fluctuations.
Infrared (IR) physics at the electroweak scale imposes a second constraint on the UV fixed-point structure: We demand that an observationally viable dynamics arises along an RG trajectory emanating from the fixed point. Whenever quantum fluctuations of gravity generate a negative scaling dimension for a matter coupling, e.g., a Yukawa coupling, they force it to a specific value at the Planck scale. Assuming no new physics, its further RG flow is given by well-known Standard Model running down to the electroweak scale. This maps a microscopic fixed-point value to an IR quantity. The fixed-point value depends on the gravitational dynamics parameterized by a set of microscopic couplings. Observational viability thus constrains the microscopic gravitational dynamics. This holds for an asymptotically safe fixed-point and in an effective field theoretic description of gravity, in principle restricting larger classes of quantum-gravity models.

In Sec.~\ref{sec:FPscenarios} we clarify the  general gravity-matter fixed-point structure in asymptotic safety, highlighting the role of global symmetries. In Sec.~\ref{sec:RG-flow} we obtain the RG-flow of a gravity-Higgs-Yukawa model. In Sec.~\ref{sec:results}, we discuss the resulting bounds on gravitational coupling space, i.e., the  weak-gravity bound (cf. Sec.~\ref{sec:weak_gravity}) and a constraint resulting from phenomenological viability at low scales (cf. Sec.~\ref{sec:phenoViability}). In Sec.~\ref{sec:effective} we explore similar bounds in an effective field theory setting beyond the realm of asymptotic safety. We conclude in Sec.~\ref{sec:conclusions}. 

\section{Quantum-gravity induced UV completion for matter}
\label{sec:FPscenarios}
Compelling hints for a quantum-gravity induced UV completion of (subsectors of) the Standard Model within the asymptotic safety paradigm exist \cite{Daum:2009dn,Benedetti:2009gn,Narain:2009fy,Narain:2009gb,Zanusso:2009bs,Vacca:2010mj,Harst:2011zx,Eichhorn:2011pc,Folkerts:2011jz,Eichhorn:2011ec,Eichhorn:2012va,Dona:2012am,Eichhorn:2013ug,Henz:2013oxa,Dona:2013qba,Dona:2014pla,Meibohm:2015twa,Percacci:2015wwa,Oda:2015sma,Dona:2015tnf,Meibohm:2016mkp,Eichhorn:2016esv,Eichhorn:2016vvy,Christiansen:2017gtg,Biemans:2017zca,Hamada:2017rvn,Falls:2017cze}
based on truncated functional RG flows. Quantum-gravity fluctuations become strong near the Planck scale and significantly alter the RG running of matter couplings. If the gravitational couplings approach an interacting fixed point, fixed-point scaling can also be induced in the matter sector. 
Here, we elucidate the induced fixed-point structure that is shaped by global symmetries of the kinetic terms of the respective matter content.

\subsection{Scenario A: Maximally symmetric asymptotic safety}
\label{sec:minimally_interacting}

\begin{figure}[!t]
\includegraphics[width=\linewidth]{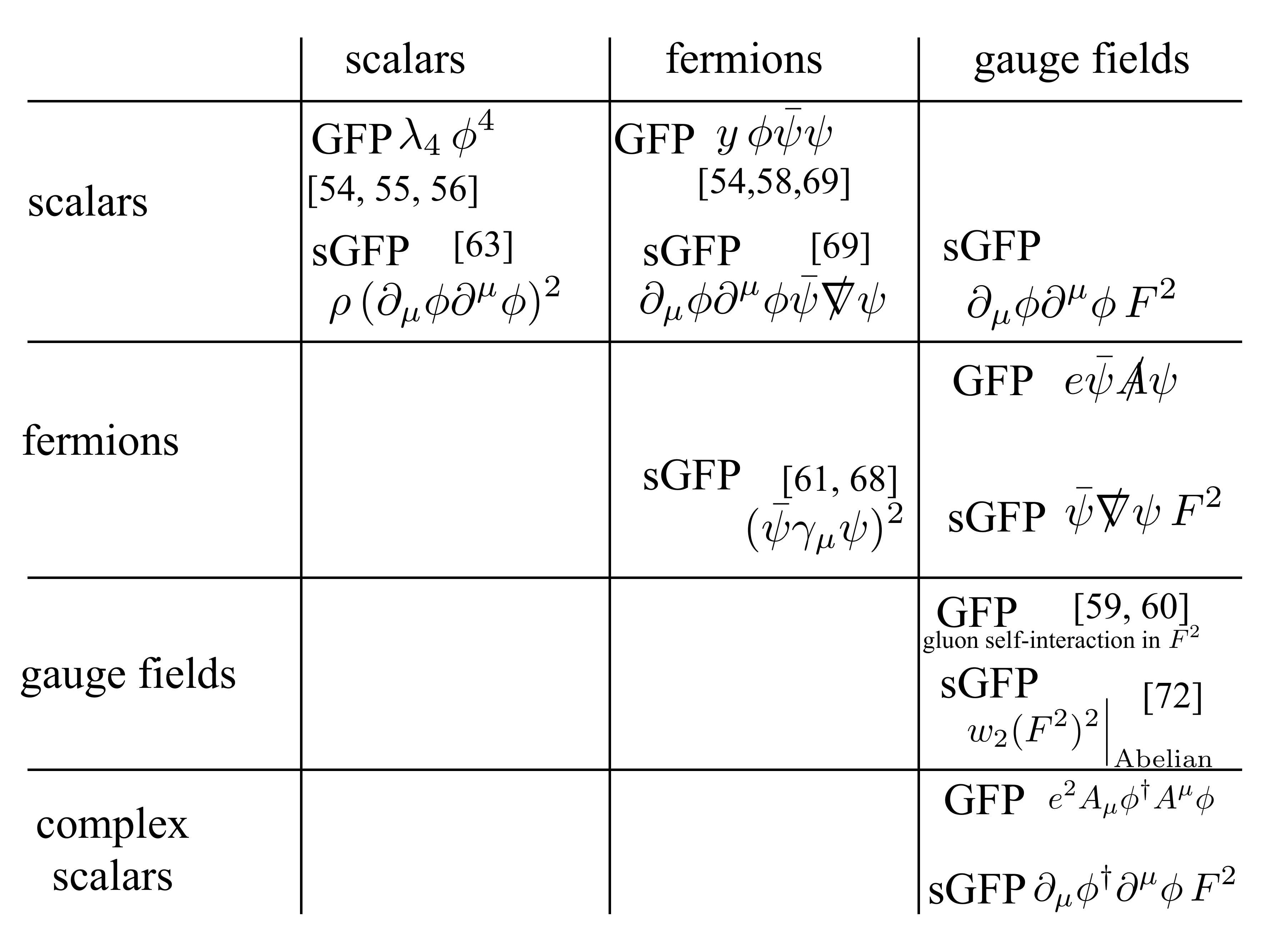}
\caption{\label{fig:table} We show interactions of scalars (complex and real), fermions and gauge fields, and point out which interactions feature a free fixed point for symmetry reasons. We also highlight examples of interactions that can only exhibit a shifted Gau\ss{}ian fixed point. For the latter, a real fixed-point value is not guaranteed, which provides us with a weak-gravity bound. Where these interactions have been explored and our symmetry-argument is supported by calculations, we indicate the corresponding references.
}
\end{figure}

\begin{figure}[!t]
\includegraphics[width=0.9\linewidth,clip=true,trim=0cm 0cm 5cm 2cm]{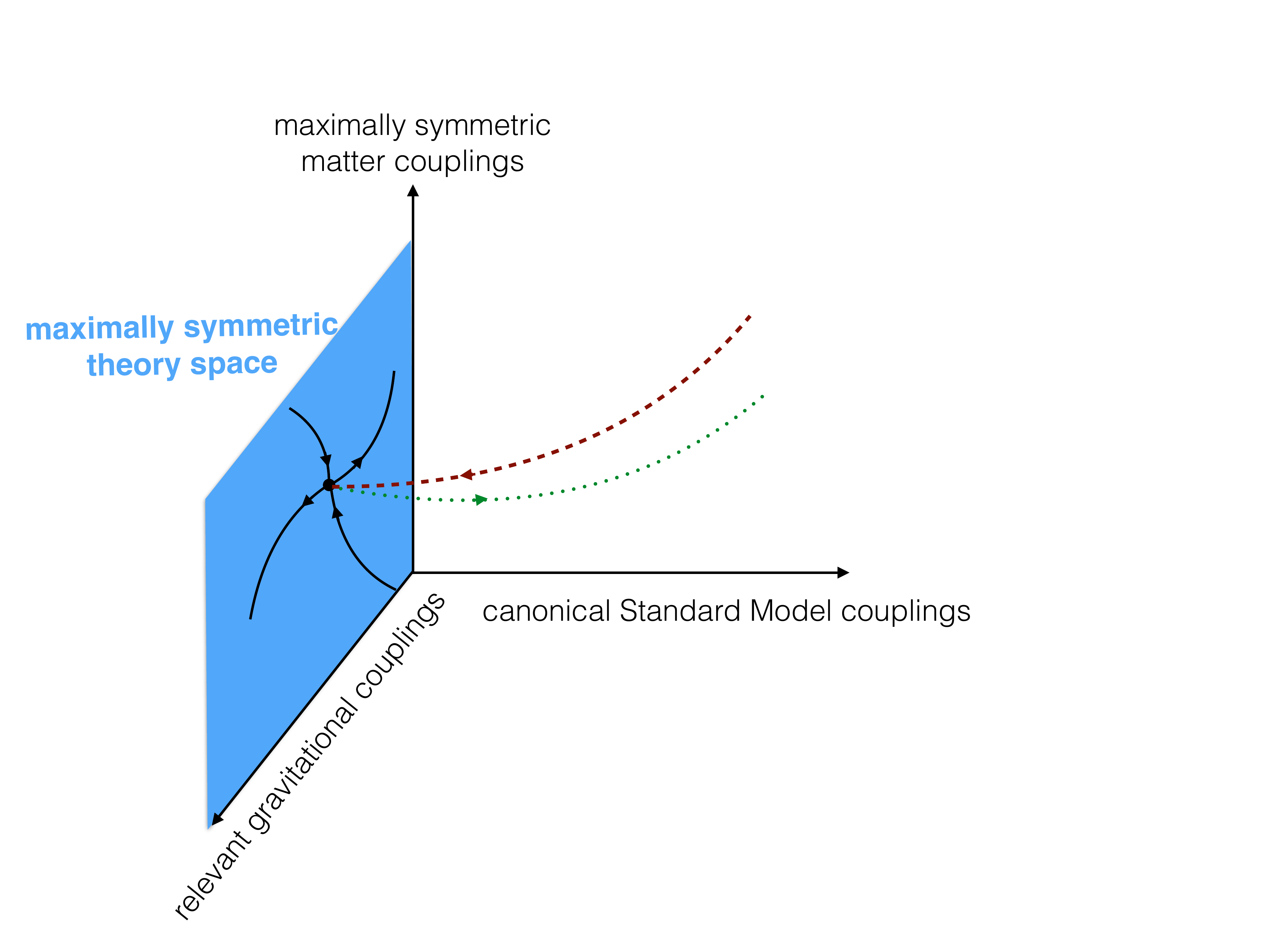}
\caption{\label{fig:ill_theoryspace} We sketch the theory space of maximally symmetric asymptotic safety: The blue (shaded) hypersurface features maximal symmetry, where all interactions respect the global symmetries of the kinetic terms. The quantum-gravity induced fixed point lies at finite values of the corresponding couplings.
Standard Model interactions lie outside this hypersurface, as they break the corresponding global symmetries. 
}
\end{figure}

In the far UV, where gravity is fully interacting, quantum gravity fluctuations induce residual matter interactions \cite{Eichhorn:2011pc,Eichhorn:2012va,Eichhorn:2013ug,Meibohm:2016mkp,Eichhorn:2016esv,Christiansen:2017gtg}. Gravity-matter systems that exhibit quantum scale-invariance necessarily feature matter self-interactions within the corresponding truncations. Thus, there is no fixed point with a fully Gau\ss ian matter sector.
The underlying mechanism is simple: As all forms of energy gravitate, quantum gravity fluctuations couple to the kinetic terms for matter fields and prevent free matter theories from becoming scale-invariant. Technically, this is a consequence of gravity loop-diagrams unavoidably generating matter self-interactions. 
A useful analogy from low-energy physics is given by the Euler-Heisenberg Lagrangian for QED, where quantum fluctuations of charged matter generate photon self-interactions, encoded in higher-order terms for photons. The pivotal difference to asymptotically safe gravity is that in the latter the interactions are \emph{not} induced by the RG flow towards the IR, but must already exist in the UV.  As a result, no matter theory can become completely asymptotically free in the presence of fixed-point gravity. A minimal set of interactions is necessarily present. The question is: Which are the residual interactions? At a first glance, it might be surprising that these are \emph{not} the canonical interactions of the Standard Model. 
 The latter vanish at the maximally-symmetric asymptotically safe fixed point, as they break the global symmetries of the kinetic terms, cf.~Fig.~\ref{fig:table}.
Instead, the induced interactions are invariant under the symmetries of the kinetic terms of the respective matter fields, since asymptotically safe gravity preserves global symmetries -- at least in all truncated RG flow that have been explored until now. All interactions that explicitly break these global symmetries can consistently be set to zero at the fixed point, cf.~Fig.~\ref{fig:ill_theoryspace}.
This does of course \emph{not} exclude the existence of a fixed point with a lower degree of symmetry, cf.~Sec.~\ref{sec:more_interacting}. We will now examine the maximally-symmetric asymptotically safe fixed point in detail and explain which interactions are and are not present.\\
Specifically, the kinetic term for scalars features a $\mathbb{Z}_2$ symmetry,
$\phi \rightarrow -\phi$, as well as shift symmetry, $\phi \rightarrow \phi+a$. These restrict the allowed interactions to a set of derivative interactions with even powers of $\phi$ \cite{Eichhorn:2012va}. This excludes a scalar potential, Yukawa interactions and a minimal coupling to gauge fields. Generalizing to a complex SU(2) scalar, all canonical interactions that are present for the Higgs in the Standard Model have to vanish at the maximally symmetric fixed point.
\\
For gauge fields, the kinetic term features a reflection symmetry, $A_{\mu} \rightarrow -A_{\mu}$, restricting the allowed interactions to those with even numbers of gauge fields. This is sufficient to conclude that non-Abelian self-interactions and the minimal gauge coupling to matter are switched off at the maximally symmetric fixed point, see \cite{Daum:2009dn, Folkerts:2011jz, Harst:2011zx, Christiansen:2017gtg}. Instead, scale-invariance appears to require photon self-interactions of the form $(F^2)^2$ which are induced by gravity \cite{Christiansen:2017gtg} in the Abelian case. We conjecture that this holds in the non-Abelian case, but that it is only the \emph{Abelian} part that will be induced, i.e., quantum gravity at the maximally symmetric fixed point ``Abelianizes" non-Abelian gauge fields. At this fixed point all interactions of non-Abelian gauge fields with matter and all self-interactions are switched off.
\\
Finally, fermions feature a global U(1) phase rotation as well as a chiral phase rotation for the kinetic term. These two can alternatively be rewritten as a separate phase rotation for left- and right-handed fermions. This precludes the coupling to the Higgs via a Yukawa term, and a minimal coupling to gauge fields is forbidden by the reflection symmetry in the gauge sector. On the other hand, specific four-fermion interactions are compatible with all symmetries and thus exist at the maximally-symmetric asymptotically safe fixed point.
\\
To summarize, quantum scale-invariance that includes quantum gravity appears to differ from classical scale invariance in a critical way (besides the general technical challenge of probing scale invariance in a framework with a mass-like regularization, see Sec.~\ref{sec:RG-flow} for details): Interactions which are canonically higher order and therefore not classically scale invariant are a necessary part of a fixed-point action in gravity-matter systems, i.e., they are required by quantum scale-invariance. Thus, the free, Gau\ss{}ian fixed point that is typically present in these couplings in a pure matter system gets shifted and becomes interacting.
Hence, quantum gravity can induce a form of ``interacting asymptotic freedom" for the Standard Model, where  a set of higher-order couplings invariant under the maximal global symmetries stay finite, while the canonically marginal couplings which break these symmetries become free, cf.~FIG.~\ref{fig:ill_theoryspace}.
\\
There are general arguments suggesting that global symmetries might be violated by quantum gravitational effects \cite{Susskind:1995da,Kallosh:1995hi}. The effect appears to be tied to the existence of black hole remnants, which is an issue that remains to be investigated further in asymptotically safe gravity, for first studies see \cite{Falls:2010he}. Moreover, the statement that a symmetry is violated is of course insufficient: Even if it is violated, the effect might be tiny; and thus it is crucial to quantify the violation of global symmetries -- if at all present -- in a given quantum gravity model.
\\

\begin{figure}[t]
	\centering
	\includegraphics[width=0.75\linewidth, clip=true, trim=0cm 2cm 2cm 1cm]{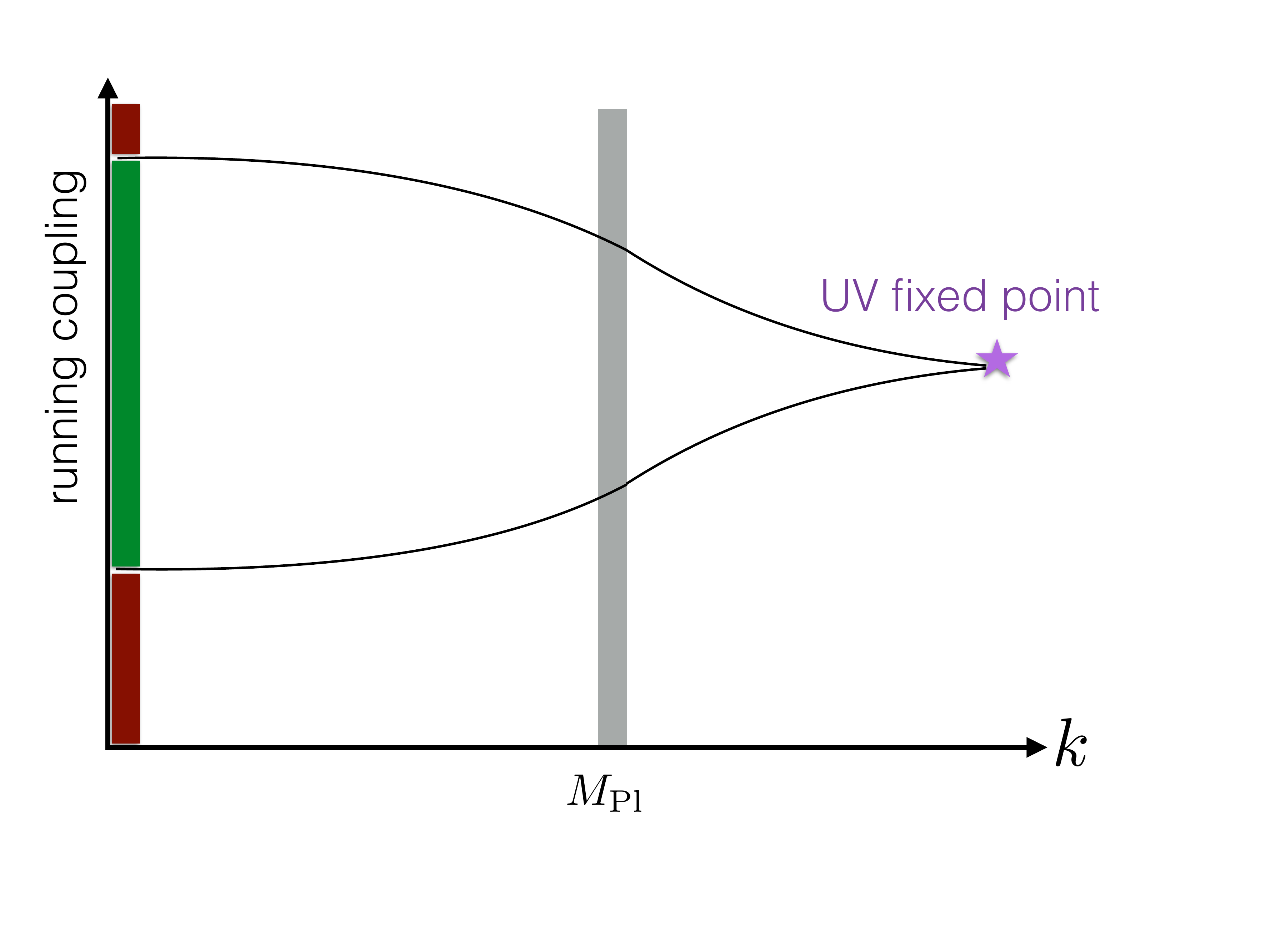}\\
	\includegraphics[width=0.75\linewidth, clip=true, trim=0cm 2cm 2cm 1cm]{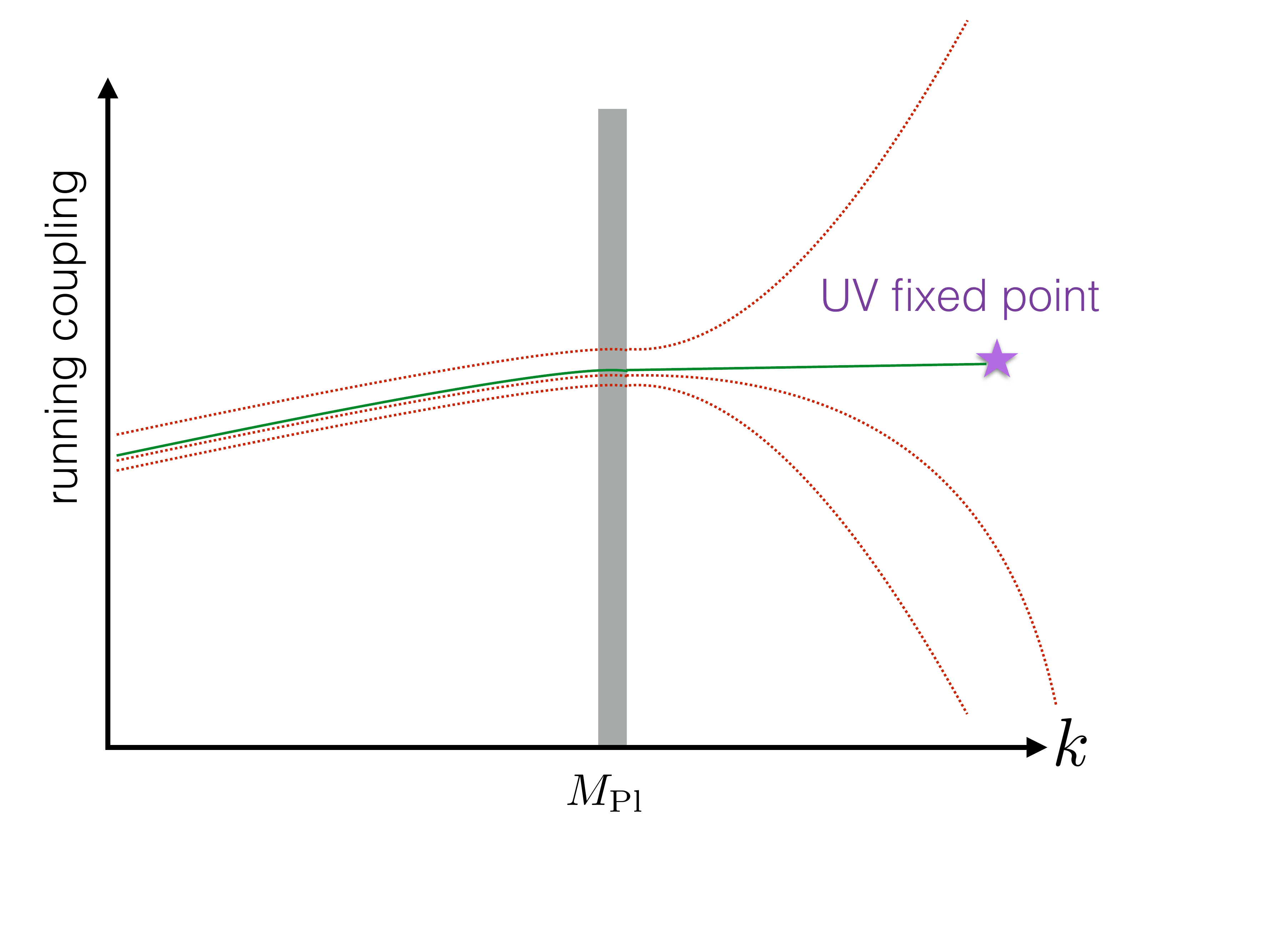}
	\caption{
	\label{fig:relevant_irrelevant}
	Schematic flow of an (ir)relevant (lower/upper panel) coupling. For a relevant coupling, the fixed point is IR-repulsive, thus a large range of IR values of the coupling can be reached starting from the fixed point (green region), whereas only those IR values outside the basin of attraction of the fixed point are excluded (dark red region). For an irrelevant coupling, quantum-gravity fluctuations force the coupling to remain at its fixed-point value down to the Planck scale. Once quantum gravity has switched off, matter fluctuations can drive the coupling away from its fixed-point value. As only one particular value can be realized at the Planck scale, the IR value of the coupling is predicted. Other IR values cannot be connected to a UV-safe regime (cf.~red curves).
	}
\end{figure}

We now discuss two observational consistency tests for the maximally-symmetric asymptotically safe fixed point.
The first is the requirement of real fixed-point values for the residual interactions: If quantum- gravity fluctuations are too strong, they can shift the fixed point  into the complex plane where it ceases to be viable. The requirement of real couplings restricts the allowed gravitational coupling space. In particular, it requires quantum gravity fluctuations to stay sufficiently weak.
\\
To recover the Standard Model at low scales, the flow towards the IR must leave the maximally-symmetric hypersurface in theory space. The second constraint arises as the maximally symmetric fixed point must be connected to observable physics in the IR by a viable RG trajectory. This test is particularly strong under the assumption of no (relevant) new physics between the electroweak and the Planck scale. In that case, the RG flow emanating from the maximally symmetric fixed point must yield the correct values of the Standard Model couplings at the Planck scale. Otherwise, the low-energy values of couplings and masses in the Standard Model do not match the observational values. Starting from the maximally-symmetric asymptotically safe fixed point the canonical Standard Model couplings remain zero all the way down to the Planck scale if quantum fluctuations render them irrelevant, cf.~Fig.~\ref{fig:relevant_irrelevant}. 
This case becomes particularly interesting for the quartic Higgs coupling, where the Planck-scale value is actually rather close to zero \cite{Degrassi:2012ry,Buttazzo:2013uya,Bezrukov:2012sa}. If quantum gravity effects render the quartic coupling irrelevant, the Higgs mass becomes predictable \cite{Shaposhnikov:2009pv,Bezrukov:2012sa}. On the other hand, if the Yukawa couplings $y$ are irrelevant at the maximally symmetric fixed point, the RG flow emanating from it would yield 
fermion masses significantly smaller than the experimentally observed values. This is particularly obvious in the case of the top mass of $\sim 176$~GeV, which requires a Planck-mass value $y\gg 0$.
\subsection{Scenario B: Asymptotic safety with Standard-Model symmetries}
\label{sec:more_interacting}
\begin{figure}[!t]
\includegraphics[width=\linewidth,clip=true,trim=0cm 0cm 5cm 2cm]{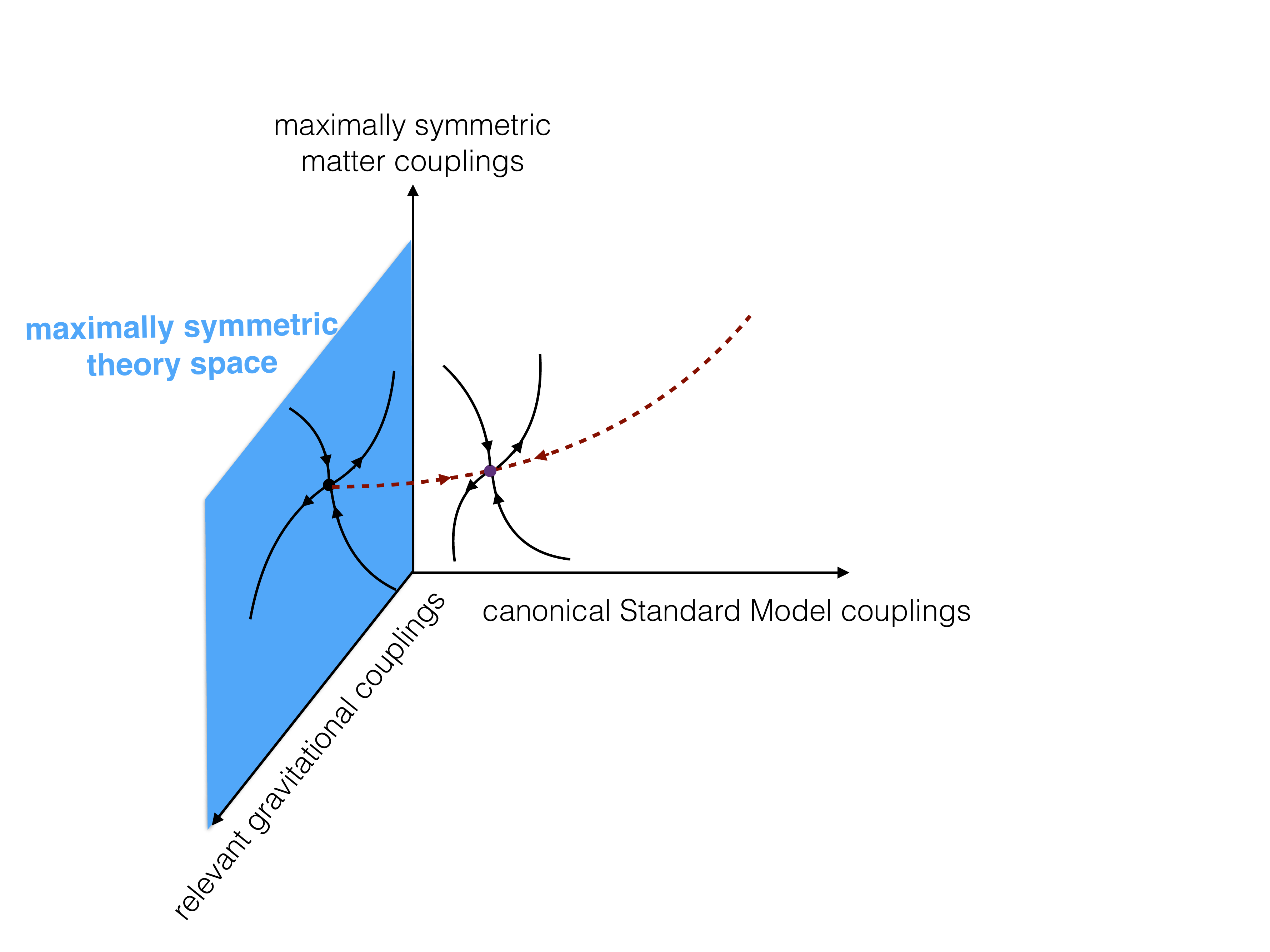}
\caption{\label{fig:ill_theoryspaceB} In the enlarged theory space which includes the canonical Standard Model couplings, quantum-gravity effects could induce a second fixed point at a finite value of the Standard Model couplings. For all Standard Model couplings that are irrelevant at that fixed point, the low-energy value is predicted.}
\end{figure}
\begin{figure}[t]
\includegraphics[width=0.9\linewidth]{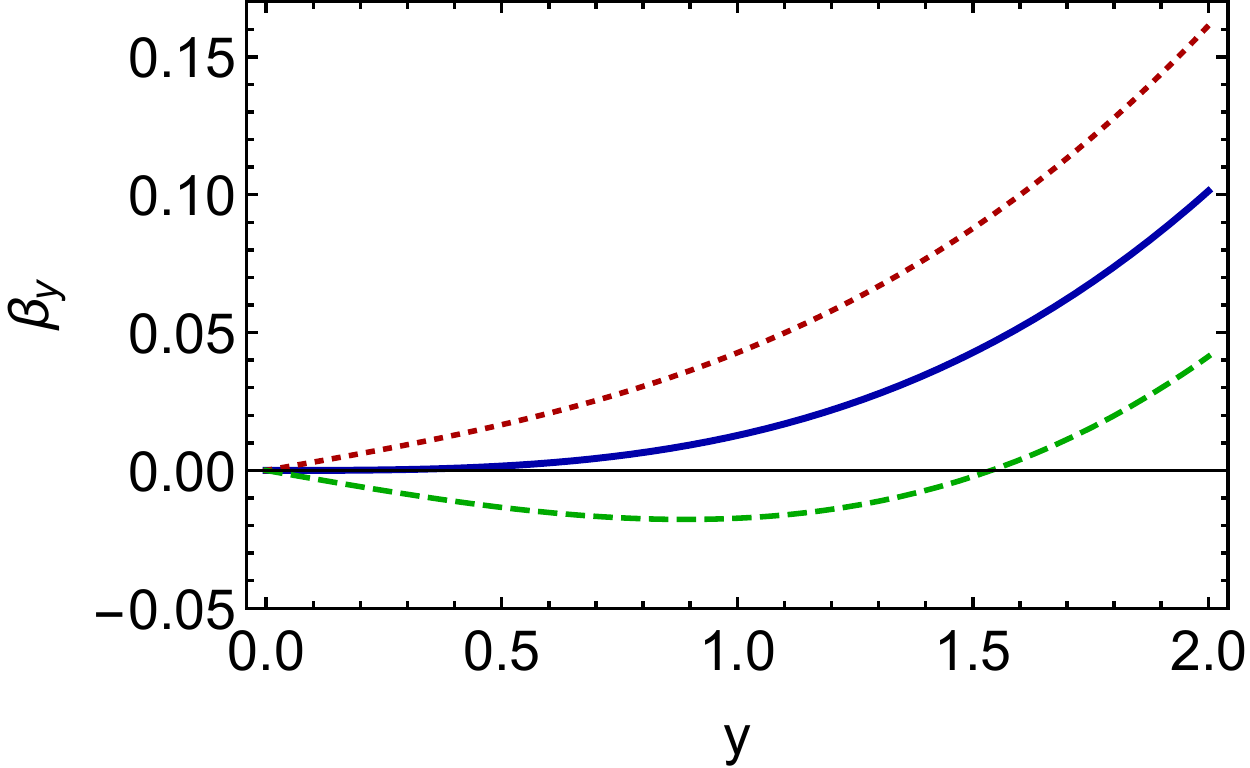}
\caption{\label{fig:betay} We show the beta function for the Yukawa coupling cf.~Eq.~\eqref{eq:betayschem}, for the case $\mathcal{D}_y>0$ (dotted  
 red line), $\mathcal{D}_y=0$ (pure matter case, continuous 
 blue line) and $\mathcal{D}_y<0$ (dashed 
 green line). For $\mathcal{D}_y<0$, $y$ becomes relevant at the free fixed point and irrelevant at a new interacting fixed point.}
\end{figure}

In addition to the maximally-symmetric asymptotically safe fixed point, interacting fixed points with explicitly broken global symmetries could exist.
For instance in a simple scalar-fermion model, breaking shift symmetry for the scalar and additionally imposing only a discrete chiral symmetry $\phi \rightarrow- \phi, \psi \rightarrow e^{i \gamma_5 \pi/2}\psi, \bar{\psi} \rightarrow e^{i \gamma_5 \pi/2} \bar{\psi}$ allows to introduce a Yukawa coupling. Within the pure matter model, its beta function features a cubic term with a positive coefficient. The quantum gravity contribution must be linear in the Yukawa coupling due to symmetry, reading
\be
\beta_y = \frac{y^3}{8 \pi^2} +  \mathcal{D}_y\; g\, y.\label{eq:betayschem}
\ee
Quantum-gravity corrections could render the fixed point at $y^\ast=0$ UV repulsive ($\mathcal{D}_y>0$). This automatically implies that no further zero of the beta function exists. If on the other hand the free fixed point is UV attractive ($\mathcal{D}_ y<0$) it is presumably compatible with IR physics. $\mathcal{D}_y<0$ simultaneously introduces another real zero (cf.~Fig.~\ref{fig:betay}) corresponding to a UV-repulsive interacting fixed point. Starting from the interacting fixed point, the Planck-scale value of $y$ is fixed, cf.~lower panel of Fig.~\ref{fig:relevant_irrelevant}. Hence, this scenario could potentially provide a way to predict the value of those Standard-Model couplings that are significantly different from zero at the Planck scale, thereby exceeding the predictive power of a perturbative setting. To give an example, the observed top mass requires a non-zero top-Yukawa coupling at the Planck scale.

The above pattern generalizes to all marginally irrelevant couplings:
The leading quantum-gravity contribution is linear in the coupling, acting like a contribution to the effective scaling dimension. For a reduced scaling dimension, quantum gravity renders the coupling relevant at the free fixed point. Simultaneously an interacting fixed point is induced. This fixed point is IR attractive, cf.~Fig.~\ref{fig:betay}. Hence, it results in a fixed finite coupling at the Planck scale, cf.~Fig.~\ref{fig:relevant_irrelevant}. This scenario places the most stringent constraints on the gravitational coupling space: The Planck-scale values of irrelevant matter couplings equal their fixed-point values which in turn are
 functions of the gravitational couplings.  Requiring an observationally viable fixed-point value for the matter coupling thus provides one additional equation that the gravitational couplings have to fulfill in addition to their fixed-point equations. Every canonically marginal coupling that is rendered irrelevant 
selects one hypersurface in the space of gravity couplings, e.g., $y=y^\ast$ in Eq.~\eqref{eq:betayschem} requires $g^{\ast}=(y^{\ast})^2/(8\pi^2 \mathcal{D}_y)$. It is a highly intriguing question whether these hypersurfaces intersect and whether the gravitational fixed-point values coincide with these intersections.

In this paper, we explore in detail how both of these phenomenologically derived tests provide strong constraints on the gravitational coupling space, by investigating a toy model of the Higgs-Yukawa sector. Those constraints arise within a truncation of the RG flow, but we conjecture that extending the truncation will only lead to quantitative changes, and not lift the existence of the constraints.

\section{RG flow for gravity and a simple Higgs-Yukawa model}
\label{sec:RG-flow}

We study  a model containing a real scalar $\phi$ and a Dirac fermion $\psi$. Maximally symmetric asymptotic safety preserves the following four global symmetries of 
the
kinetic terms:
\begin{itemize}
	\item $\mathds{Z}_2$ symmetry for the scalar: $(\phi\rightarrow -\phi)$
	\item scalar shift symmetry: $\phi \rightarrow \phi + a$
	\item U(1) phase: $\psi \rightarrow e^{i \alpha'}\psi$ and $\bar{\psi} \rightarrow \bar{\psi}e^{i\alpha'}$
	\item chiral U(1): $\psi \rightarrow e^{i \gamma_5 \alpha} \psi$ and $\bar{\psi} \rightarrow \bar{\psi}e^{i\gamma_5 \alpha}$
\end{itemize}
The maximally symmetric theory space contains higher-order interaction terms of which we include the leading-order fermionic \cite{Eichhorn:2012va,Meibohm:2016mkp} and mixed (fermion-scalar) \cite{Eichhorn:2016esv} ones. We neglect scalar interactions that exist in the maximally symmetric theory space \cite{Eichhorn:2012va, Eichhorn:2013ug}, since the present truncation is sufficient to illustrate the fixed-point structure and both constraints discussed in Sec.~\ref{sec:FPscenarios}.
Our truncation for the scale-dependent effective action in the matter sector with maximal global symmetry is
\bea
\Gamma_{k,\text{MS}} &=& \frac{Z_{\phi}}{2}\int
d^4x \sqrt{g} \left(g^{\mu \nu} \partial_{\mu}\phi \partial_{\nu}\phi
  \right)\label{eq:effActionMS} \\
&+& iZ_{\psi} \int d^4 x \sqrt{g} \bar{\psi}\slashed{\nabla} \psi
\nonumber\\
&+& \frac{1}{2} \int d^4x \sqrt{g} \Big[\bar{\lambda}_A \, (i\bar\psi\gamma_\mu\gamma_5\psi)(i\bar\psi\gamma^\mu\gamma_5\psi)
\nonumber\\
&{}&\quad\quad\quad\quad\quad\quad + \bar{\lambda}_V \, (\bar\psi\gamma_\mu\psi)(\bar\psi\gamma^\mu\psi) \Big]
\nonumber\\
&+&\int d^4x\sqrt{g}\Big[i \bar{\chi_1}\left(\bar\psi\gamma^\mu\nabla_\nu\psi -
    (\nabla_\nu\bar\psi)\gamma^\mu\psi\right)
  \left(\partial_\mu\phi\partial^\nu\phi\right)
\nonumber\\
&{}&\quad\quad\quad\quad	+  i\bar{\chi}_2\left(\bar\psi\gamma^\mu\nabla_\mu\psi -
    (\nabla_\mu\bar\psi)\gamma^\mu\psi\right)
  \left(\partial_\nu\phi\partial^\nu\phi\right)
  \Big].
  \notag
\eea
In the presence of quantum gravity $\bar{\lambda}_A$ is finite
\footnote{
	For the four-fermion interactions, there are two independent ones that are symmetric under the separate left- and right handed phase rotations. However, gravity only induces one of them. In \cite{Eichhorn:2011pc}, this becomes clear when one uses a parameterization in terms of $\lambda_A =1/2(\lambda_+ - \lambda_-)$ and $\lambda_V=1/2 (\lambda_++\lambda_-)$; and in \cite{Meibohm:2016mkp} one can see this by taking into account the Fierz identity that allows one to make the transition to $\lambda_A$ and $\lambda_V$, and then observing that the gravity-induced contribution to the beta functions respects $\beta_{\lambda_{\sigma}}\Big|_{\rm induced} = -2\beta_{\lambda_V}\Big|_{\rm induced}$.
	It is curious that the four-fermion interactions which are induced by gauge interactions do \emph{not} follow this pattern, i.e., there are two independent four-fermion interactions that are induced \cite{Gies:2003dp,Gies:2005as,Braun:2006jd}. Technically, the reason lies in the different diagram that induces four-fermion interactions in gravity as compared to gauge theories, see \cite{Eichhorn:2011pc}, and the different structure of the corresponding vertex. This structure holds for all choices of the gauge parameter $\beta$, as the two-graviton-fermion-antifermion-vertex vanishes for all scalar components of the graviton. Moreover, the vertex also vanishes for the transverse vector component of the graviton, therefore the result also holds for all values of $\alpha$.
}
\cite{Eichhorn:2012va,Meibohm:2016mkp} and $i\bar{\chi}_{1/2}\neq 0$ as well \cite{Eichhorn:2016esv}. Our work is the first to combine both induced interaction channels and include the resulting anomalous dimensions. In terms of physics, we mainly aim at connecting the maximally-symmetric asymptotically safe fixed-point structure to the top-quark mass. Within the toy model at hand, the low-energy fermion mass is potentially impacted by four-fermion interactions and by fermion-scalar interactions through the anomalous dimensions.

Just as the top-Yukawa coupling is the source of the top-quark mass in the Standard Model a simple Yukawa coupling would generate an IR fermion mass after spontaneous symmetry breaking in the scalar sector. The Yukawa coupling is \emph{not} part of the maximally-symmetric theory space as it preserves only a discrete chiral symmetry under which $\phi\rightarrow-\phi$, $\psi \rightarrow e^{i \frac{\pi}{2}\gamma_5} \psi$ and $\bar{\psi} \rightarrow \bar{\psi}e^{i\frac{\pi}{2}\gamma_5}$. The flowing action containing the leading-order operators departing from the maximally symmetric hypersurface (cf.~Fig.~\ref{fig:ill_theoryspaceB}) reads
 \bea
\Gamma_{k\, \rm SM}&=& \int d^4x\sqrt{g}\, \Bigl( iy\phi\bar\psi\psi +\bar{\lambda}_S(\bar\psi\psi)(\bar\psi\psi)\Bigr).
 \eea
To elucidate the impact of higher-order couplings in scenario B (cf.~Sec.~\ref{sec:more_interacting}), we also include an additional four-fermion interaction, parameterized by $\bar\lambda_S$. Together with $\bar\lambda_A$ and $\bar\lambda_V$ it constitutes a Fierz-complete basis of interactions invariant under the discrete chiral symmetry. To search for a scale-invariant fixed point regime we make a transition to dimensionless couplings
\bea
\lambda_X = \frac{\bar{\lambda}_X}{Z_{\psi}^2} k^2, \quad \chi_i =\frac{\bar{\chi}_i}{Z_\psi\, Z_{\phi}} k^4
\eea
and introduce the anomalous dimensions
\be
\eta_{\phi} = -\partial_t \ln Z_{\phi} = -k\partial_k \ln Z_{\phi}, \quad \eta_{\psi} = - \partial_t \ln Z_{\psi}\;.
\ee

\subsection{Wetterich equation}
\begin{figure}[t]
	\centering
	\includegraphics[width=\linewidth]{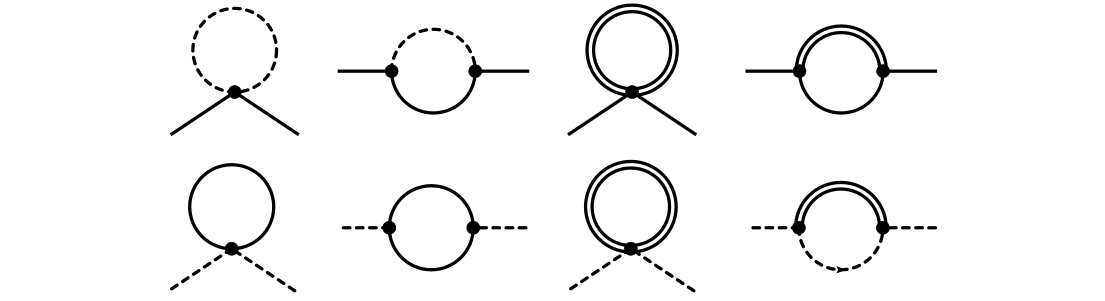}
	\caption{
	\label{fig:flowDiagsAnomDim}
 Flow diagrams of $\chi$-, Yukawa- and gravitational contributions to the anomalous dimensions of fermions (cf. Eq.~\eqref{eq:etapsi}) and scalars (cf. Eq.~\eqref{eq:etaphi}). Metric fluctuations, fermions and scalars are represented by double, 
solid and dashed lines respectively.}
\end{figure}

\begin{figure}[t]
	\centering
	\includegraphics[width=\linewidth]{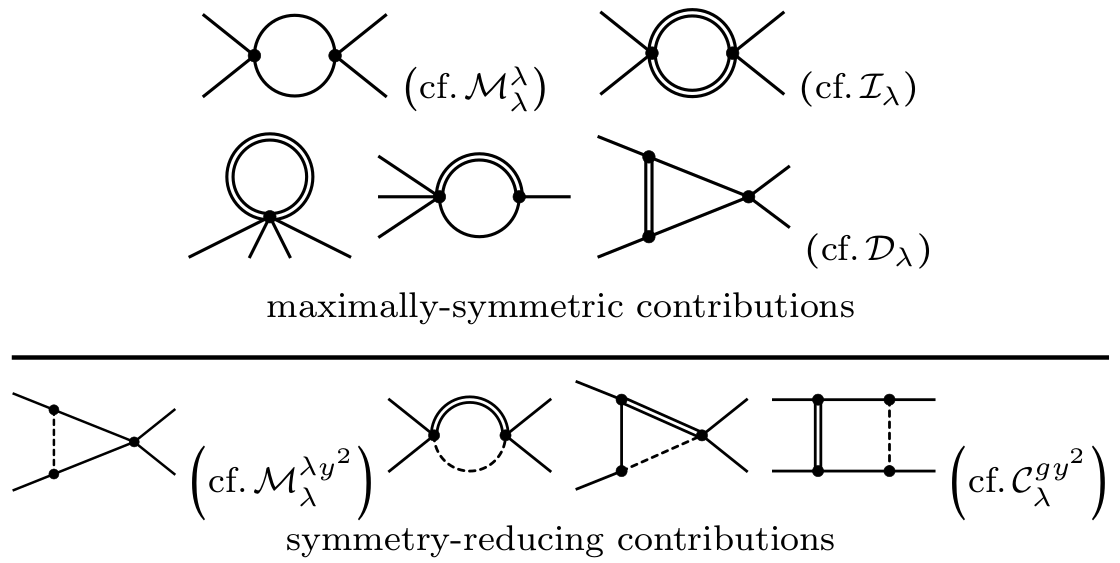}
	\caption{
	\label{fig:flowDiags4fermi}
	Diagrams contributing to the maximally symmetric flows of four-fermion interactions: the pure-matter (cf. $\mathcal{M}_\lambda^\lambda$ in Eq.~\eqref{eq:beta_lamda_A},\eqref{eq:beta_lamda_V} and \eqref{eq:beta_lamda_S}) and the gravity-induced contributions in the first line (cf. $\mathcal{I}_\lambda$ in Eq.~\eqref{eq:beta_lamda_A}); gravity-contributions to anomalous scaling in the second line (cf. $\mathcal{D}_\lambda$ in Eq.~\eqref{eq:beta_lamda_A}, \eqref{eq:beta_lamda_V} and \eqref{eq:beta_lamda_S}). 
\\
The symmetry-reducing diagrams in the last line are absent for vanishing Yukawa coupling (cf. $\mathcal{M}_\lambda^{\lambda y^2}$ in Eq.~\eqref{eq:beta_lamda_A}, \eqref{eq:beta_lamda_V} and \eqref{eq:beta_lamda_S} and $\mathcal{C}_\lambda^{gy^2}$ in Eq.~\eqref{eq:beta_lamda_S}). There are further diagrams that vanish.
}
\end{figure}

\begin{figure}[t]
	\centering
	\includegraphics[width=\linewidth]{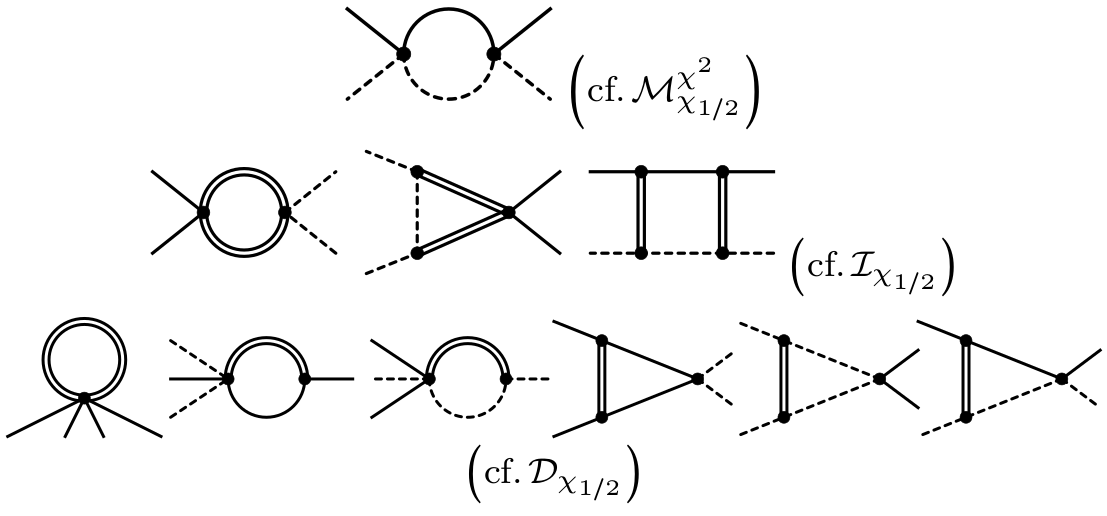}
	\caption{
	\label{fig:flowDiagsChi}
 Diagrams contributing to the maximally symmetric flow of two-fermion--two-scalar interactions: the pure-matter contributions in the first line (cf. $\mathcal{M}_{\chi_{1/2}}^{\chi^2}$ in Eq.~\eqref{eq:beta_Chi1} \& \eqref{eq:beta_Chi2}, while the second diagram vanishes decoupling $\lambda_{A/S/V}$ from $\beta_{\chi_{1/2}}$); the gravity-induced contributions in the second line (cf. $\mathcal{I}_{\chi_1/2}$ in Eq.~\eqref{eq:beta_Chi1} \& \eqref{eq:beta_Chi2}); the gravity-
contributions to anomalous scaling in the third line (cf. $\mathcal{D}_{\chi_1/2}$ in Eq.~\eqref{eq:beta_Chi1} \& \eqref{eq:beta_Chi2}). There are further diagrams that vanish and we neglect contributions in the symmetry-reducing flows (i.e., for $y\neq 0$) since we do not calculate in this regime.
 }
\end{figure}

\begin{figure}[t]
	\centering
	\includegraphics[width=\linewidth]{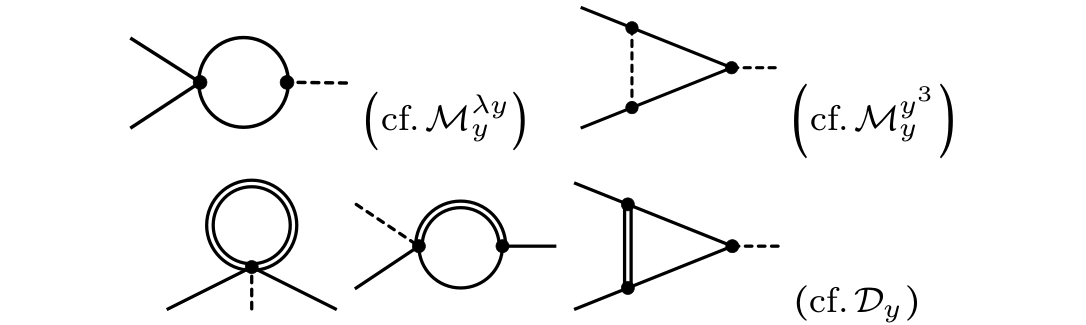}
	\caption{
	\label{fig:flowDiagsYukawa}
	 Diagrams contributing to the flow of the Yukawa interaction: pure-matter contributions in the first line (cf. $\mathcal{M}_{y}^{\lambda y}$ and $\mathcal{M}_{y}^{y^3}$ in Eq.~\eqref{eq:beta_y}); gravity-
 contributions to anomalous 
 scaling in the second line (cf. $\mathcal{D}_y$ in Eq.~\eqref{eq:beta_y}). There are further diagrams that vanish.
	}
\end{figure}

To study the scale dependence and extract the $\beta$-functions of the couplings we employ the functional RG. It relies on a mass-like regulator $R_k(p^2)$ in the path integral that suppresses quantum fluctuations below $k$, while higher ones are integrated out. This provides a flowing action, which yields the quantum equations of motion with the effect of all high-momentum modes included, while the low-momentum ones are suppressed. The change of the flowing action under a change of $k$ is driven by modes with momenta $p^2 \approx k^2$  and governed by a formally exact one-loop flow equation \cite{Wetterich:1992yh}
\begin{align}
\partial_t \Gamma_k := k \partial_k \Gamma_k = \frac{1}{2} {\rm Tr}\,
\left(\Gamma_k^{(2)}+ R_k \right)^{-1} \partial_t
R_k,\label{eq:flow} 
\end{align}
see also \cite{Morris:1993qb}, for reviews see \cite{Berges:2000ew, Polonyi:2001se,
Pawlowski:2005xe, Gies:2006wv, Delamotte:2007pf, Rosten:2010vm, Braun:2011pp}.
The trace $\rm Tr$ runs over the continuous and/or discrete eigenvalues of the full, regularized field- and momentum-	dependent  propagator $\left(\Gamma_k^{(2)}+ R_k \right)^{-1}$. On a flat background, it translates into a loop-momentum integral, a summation in field space and a trace over all internal and Lorentz indices. This equation has an exact 1-loop diagrammatic representation. Expanding in external fields, a series of non-perturbative 1-loop diagrams represents the contributions to the beta function of a given coupling. For instance, Figs.~\ref{fig:flowDiagsAnomDim},~\ref{fig:flowDiags4fermi},~\ref{fig:flowDiagsChi}~and~\ref{fig:flowDiagsYukawa} 
show the contributions to the flow of the anomalous dimensions $\eta_{\phi/\psi}$, four-fermion couplings $\lambda_i$, mixed two-fermion--two-scalar couplings $\chi_{1/2}$ and the Yukawa coupling $y$, respectively within our truncation of the dynamics.
\subsection{Parameterizing the propagator of metric fluctuations}
\label{sec:gravity_truncation}
The effect of gravity on matter couplings is given in terms of the flat-space propagator of metric fluctuations $(\Gamma_k^{(hh)})^{-1}$. We  derive it from 
\begin{align}
	\label{eq:effActionGrav}
	\Gamma_{k\,\text{grav}}= 
	&\frac{-1}{16 \pi \bar{G}}Z_h \int d^4x\, \sqrt{g}\left( R- 2\bar{\lambda} \right) + S_\text{gauge-fixing}
	\\
	&+\frac{1}{16 \pi \bar{G}}Z_h \int d^4x\, \sqrt{g}
	\left( \bar{a} R^2 + \bar{b} R_{\mu\nu}R^{\mu\nu} \right.\nonumber\\
	& \quad\quad\quad\quad\quad\quad\quad\quad\left. + \bar{c} R\Box R + \bar{d}R_{\mu\nu} \Box R^{\mu\nu} \right)\;. \notag
\end{align}
The Einstein-Hilbert action in the first line of Eq.~\eqref{eq:effActionGrav} is supplemented by a complete set of terms up to sixth order in derivatives which can contribute to the graviton propagator on flat space, see \cite{Decanini:2008pr}. While the presence of higher-order terms might raise the question of unitarity -- even though unitarity can be preserved in the presence of such terms \cite{unitary_ChrisDanielFrank} -- it cannot be meaningfully addressed in finite truncations like ours \cite{Barnaby}.
We introduce the  gravity anomalous dimension 
\be
\eta_h = - \partial_t \ln Z_h,
\ee
and make a transition to dimensionless couplings
\bea
g&=& \bar{G} k^2, \quad \mu_h = -2 \bar{\lambda}k^{-2}, \quad a = \bar{a}k^2, \quad b = \bar{b} k^2,\nonumber\\
c&=& \bar{c}k^4, \quad d = \bar{d}k^4.
\eea
As the setup of an RG flow with the interpretation of a local coarse-graining procedure requires the distinction of high-energy and low-energy modes, we must introduce a background. To that end, we employ the background field method, with a linear split, such that
\be
g_{\mu \nu} = \bar{g}_{\mu \nu}+ h_{\mu \nu}\;.
\ee
Specifically, we work with the simplest choice
\be
\bar{g}_{\mu \nu} = \delta_{\mu \nu}\;.
\ee
While projections on operators including curvature invariants could differ in cases where the background is insufficient to distinguish between different invariants, projections on pure-matter couplings do not depend on the choice of background. An expansion around a flat background does therefore not introduce additional errors beyond those due to contributions from operators beyond the scope of our truncation. In fact, all explicit curvature dependence of the graviton propagator drops out of the $\beta$-functions of the pure matter couplings that we focus on, although it might affect non-minimal couplings that we do not include here.
We choose a standard background-field gauge fixing, see, e.g., Eq.~(2), (3) in \cite{Gies:2015tca} with gauge parameters $\beta=\alpha=0$, in which all but the trace and the transverse traceless (TT) modes of the York-decomposed \cite{York:1973ia} propagator of metric fluctuations decouple from the matter beta functions. A different choice of $\beta$ has been argued for in \cite{Wetterich:2016ewc}. For the regulator, we choose its tensor structure equal to that of the inverse propagator, multiplied by a scalar shape-function. We employ the Litim cutoff \cite{Litim:2001up} with arguments such that it is spectrally \cite{Gies:2002af} and RG adjusted \cite{Litim:2002xm}, i.e.,
\be
\label{eq:cutoff-spectrally}
R_{k\,\mu \nu\kappa\lambda} = \left(\Gamma_{k\, \mu \nu \kappa \lambda}^{(2)}(k^2)- \Gamma_{k\,\mu \nu \kappa \lambda}^{(2)}(p^2) \right)\theta(k^2-p^2)
\ee
for metric fluctuations. This simplifies the loop-integral on the right-hand side of the flow equation, as 
\be
\Gamma_k^{(2)}(p^2)+ R_k \rightarrow \Gamma_k^{(2)}(k^2),
\ee
i.e., there is no momentum-dependence in the denominator. For the matter fields 
\bea
R_{k\, \phi} &=& Z_{\phi} \left(k^2-p^2\right)\theta(k^2-p^2),\nonumber\\
 R_{k\, \psi} &=& Z_{\psi}\slashed{p}\left(\sqrt{\frac{k^2}{p^2}}-1 \right)\theta(k^2-p^2).
\eea
We test the regulator dependence of our results (cf.~App.~\ref{app:non-spectral}), by using a cutoff that is not spectrally adjusted, i.e., does not depend on the essential couplings.
\subsection{Spin-2-approximation}
\label{sec:TTmodeApprox}
In our off-shell, gauge-fixed setting, metric fluctuations consist of two different modes. 
A simple mode-counting argument suggests that the transverse-traceless spin-2 mode (cf.~Eq.~\eqref{eq:TTmode}) dominates the RG flow. This is particularly clear in our gauge, where the transverse vector and one of the two scalars decouple, and the flow is driven by the gauge-independent spin-2 contribution and a gauge-dependent scalar contribution.
We perform a York decomposition to find the propagator for the transverse-traceless mode $h_{\mu\nu}^{\rm TT}$ with $\partial^{\mu}h_{\mu\nu}^{\rm TT}=0,\, h^{\mu\,\, \rm TT}_{\mu}=0$ and the trace mode $h = h^{\mu}_{\mu}$:
\begin{align}
\label{eq:TTmode}
	\left(\Gamma_{k, \, \rm TT}^{(2)}(p^2)\right)^{-1} &= 
	\frac{32 \pi \bar{G}}{p^2{-2 \bar{\lambda}}+ \bar{b} p^4- \bar{d} p^6} P^{\rm TT}_{\mu\nu\kappa\lambda}(p)\;,
	\\
\label{eq:Trmode}
	\left(\Gamma_{k\, h}^{(2)}(p^2) \right)^{-1}&= -\frac{256 \pi  \bar{G}}{3 p^2{-4 \bar{\lambda}}-6 p^4 (3 \bar{a}+\bar{b})+6 p^6 (3 \bar{c}+\bar{d})}\;,
\end{align}
where the projector $P^{\rm TT}_{\mu\nu\kappa\lambda}(p)$ is given, e.g., in Eq. (B8) of \cite{Eichhorn:2010tb}.
Demanding a weak or vanishing gauge-dependence also suggests that the TT-part must dominate. 
The scalar contribution is expected to be subdominant compared to the spin-2-contribution, as it corresponds to just one instead of five off-shell modes. 
In the following, we report our results in the spin-2-approximation in the main text, which consists in neglecting the contribution of fluctuations of the trace mode.
We caution that the TT mode does not necessarily dominate in that region of coupling space where gravity is switched off dynamically, e.g., at large $\mu_h$: For our choice of regulator, the TT mode falls off with $(1+\mu_h)^{-1}$, whereas the trace mode shows a slower fall-off, $(1+2/3\,\mu_h)^{-1}$.
Our full results which include the trace mode can be found in App.~\ref{app:TT-mode-justify}.
In this work, we do not evaluate the running of the gravitational couplings in order to determine their fixed-point values within a specific truncation. Instead we treat them as free parameters. Hence, our work allows a comprehensive view on the possible microscopic gravitational dynamics constrained by observational viability.

The aforementioned suppression can be observed in Eq.~\eqref{eq:thresholds_start}-\eqref{eq:thresholds_end}, where all trace-mode threshold-integrals are significantly suppressed. This goes hand in hand with the observation that this approximation also works well in the pure-gravity sector  \cite{Eichhorn:2010tb}, and moreover fixed-point results in unimodular gravity are very similar, see, e.g., \cite{Eichhorn:2013xr,Saltas:2014cta, Eichhorn:2015bna}. For studies that highlight the fixed-point structure in the opposite approximation which keeps the conformal mode only, see, e.g., \cite{Reuter:2008wj,Reuter:2008qx,Machado:2009ph,Bonanno:2012dg,Labus:2016lkh,Dietz:2016gzg}.
\\
Our results suggest that gravity couplings that do not enter the propagator of spin-2-fluctuations are negligible for a leading-order understanding of gravity-matter dynamics. This provides guidance for the setup of truncations for the gravitational RG flow, as we provide 
input on which couplings dominate the RG flow at leading order, and which ones only add subleading corrections. 
Specifically, as only functions of the Ricci tensor, not of the Ricci scalar, enter the TT-propagator (cf.~Eq.~\eqref{eq:TTmode}), a quantitative determination of the fixed-point values of the former appears crucial.
\subsection{$\beta$-functions}
\label{sec:betas}
The $\beta$-functions of the matter couplings $y,\lambda_A,\lambda_S,\lambda_V$, $\chi_1$ and $\chi_2$ result from suitable projections \cite{Eichhorn:2011pc,Eichhorn:2016esv} of the flow equation \eqref{eq:flow} and take the form 
\begin{align}
   \label{eq:beta_Chi1}   
   \beta_{\chi_1} = &\; 
   \chi_1\left(\eta _{\psi }+\eta_\phi + 4\right)
   +\mathcal{M}_{\chi_1}^{\chi^2}
   +g^2\mathcal{I}_{\chi_1}
   +g\mathcal{D}_{\chi_1}
   \\[1ex]
    \label{eq:beta_Chi2}   
   \beta_{\chi_2} = &\; 
   \chi_2\left(\eta _{\psi }+\eta_\phi + 4\right)
   +\mathcal{M}_{\chi_2}^{\chi^2}
   +g^2\mathcal{I}_{\chi_2}
   +g\mathcal{D}_{\chi_2}
   \;,
   \\[1ex]
	\label{eq:beta_lamda_A}
   \beta_{\lambda_A} = &
   \;2 \lambda _A \left(\eta _{\psi }+1\right)
   +\frac{4}{3}\left(\lambda _A-\lambda _V\right)y^2\;
   \mathcal{M}_\lambda^{\lambda y^2}
   \notag\\&
 	+\left(\lambda _A^2-\lambda _A \left(3 \lambda _S+2 \lambda _V\right)+3 \lambda _V \left(\lambda _S-\lambda_V\right)\right)
 	\mathcal{M}_\lambda^\lambda
 	\notag\\[1ex]&
 	 +g^2\, \mathcal{I}_\lambda
   +g\lambda_A \,\mathcal{D}_\lambda
  \;,
   \\[1ex]
   \label{eq:beta_lamda_V}
   \beta_{\lambda_V} = &\;
   2\lambda _V\left(\eta _{\psi }+1\right)
 	+2\left(\lambda _A-2 \lambda _V\right)y^2\;
	\mathcal{M}_\lambda^{\lambda y^2}   
  	\notag\\[1ex]&
  	-2\left(\lambda _A \left(\lambda _S+2 \lambda _V\right)
  	+2 \lambda _V \left(\lambda _V-\lambda_S\right)\right)
  	\mathcal{M}_\lambda^\lambda
	\notag\\[1ex]&
   +g\lambda_V \,\mathcal{D}_\lambda\;,
   \\[1ex]
    \label{eq:etaphi}
   \eta_{\phi} =& \;g\,\mathcal{D}_{\eta_\phi} + \mathcal{M}_{\eta_\phi}^\chi + y^2 \mathcal{M}_{\eta_\phi}^{y^2}\;,
   \\[1ex]
   \label{eq:etapsi}
   \eta_{\psi}
   =& \;g\,\mathcal{D}_{\eta_\psi} + \mathcal{M}_{\eta_\psi}^\chi + y^2 \mathcal{M}_{\eta_\psi}^{y^2}
   \;,
   \\[1ex]
   \label{eq:beta_y}
	\beta_y = &
	\;y \left(\eta _{\psi }+\frac{\eta _{\phi }}{2}\right)
	+y^3\; \mathcal{M}_y^{y^3}
	\notag\\&
	-y\left(4 \lambda _A-3 \lambda _S+4 \lambda _V\right)\mathcal{M}_y^{\lambda y}
	+gy\, \mathcal{D}_y
   \\[1ex]
   \label{eq:beta_lamda_S}
   \beta_{\lambda_S} = &
   \;2 \lambda _S\left(\eta _{\psi }+1\right) 
   +\lambda _S \left(-6 \lambda _A+\lambda _S+2 \lambda _V\right)
   \mathcal{M}_\lambda^\lambda
   \notag\\[1ex]&
   +2\left(2 \lambda _A+\lambda _S-6 \lambda _V\right)y^2\;
   \mathcal{M}_\lambda^{\lambda y^2}
   \notag\\[1ex]&
   +g\lambda _S \,\mathcal{D}_\lambda
	+gy^2\,\mathcal{C}_\lambda^{gy^2}
	\;.
\end{align}
Observe that gravity contributions which necessitate non-zero fixed-point values are only present in those couplings respecting the maximal global symmetry. These are the inducing contributions parameterized by the threshold integrals $\mathcal{I}_i(g,\mu_h,a,b,c,d,\eta)$.
All matter beta functions feature a gravity-contribution \emph{linear} in the matter coupling itself.
These terms mimic a scaling dimension and are denoted by $\mathcal{D}_i(g,\mu_h,a,b,c,d,\eta)$.
Further, a subset of beta functions, e.g., $\beta_{\lambda_S}$, also include combined contributions from gravity and other matter couplings. These are denoted by $\mathcal{C}_i^j(g,\mu_h,a,b,c,d,\eta)$.
In $\beta_{\chi_{1/2}}$, we have already inserted the fixed-point value $y^\ast=0$.
Finally, pure matter contributions to $\beta_i$ depending on the couplings $j$ are given by $\mathcal{M}_i^j(\eta)$.
All threshold integrals are given in App.~\ref{app:thresholds}. Here, we reconfirm previous results on $\beta_{\lambda_{A,V}}$ from \cite{Eichhorn:2011pc}, and on $\beta_y$ and $\beta_{\chi_{1,2}}$ from \cite{Eichhorn:2016esv} and find agreement with \cite{Hamada:2017rvn} in the TT-mode sector (a full agreement is not expected due to different gauge choices.)
In the matter sector our results for the matter contributions to the anomalous dimensions agree with Eq.~(36)~\&~(37) in \cite{Braun:2010tt}.
\\
The given  
set of $\beta$-functions allows us to derive constraints on the gravity couplings from the structure of the matter sector.

\subsection{Gravity rules: suppressed backcoupling of induced matter interactions}
\label{sec:matterSuppression}
\begin{figure}[t]
	\centering
	\includegraphics[width=0.9\linewidth]{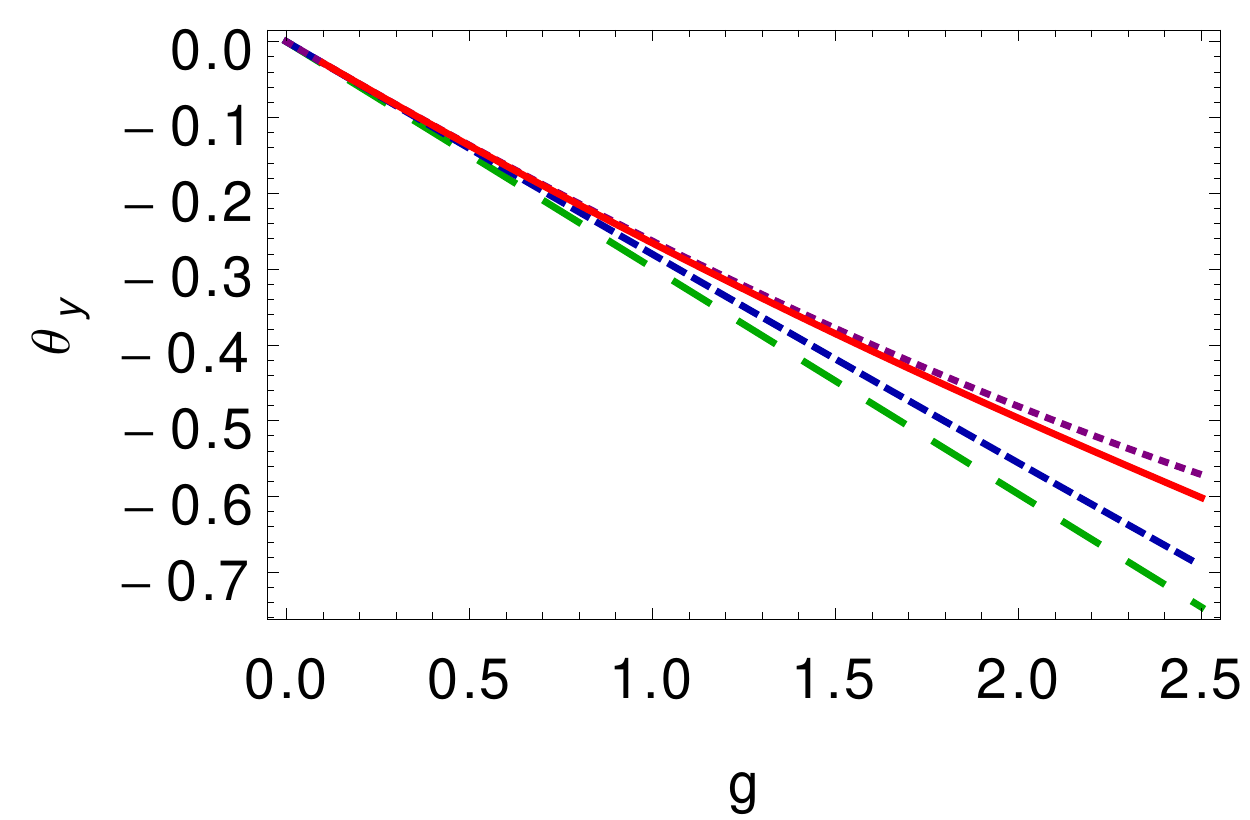}
	\caption{\label{fig:thetaY_contributions}
		Comparison of the critical exponent of the Yukawa coupling $\theta_y$ (with $a=b=c=d=\mu_h=0$) for different approximations: the green wide-dashed line shows the spin-2 contributions (including anomalous dimensions) only; the blue narrow-dashed line includes trace-contributions; the purple dotted line includes trace-contributions and matter-mediated effects from the induced four-fermion coupling $\lambda_A$;	the red continuous line additionally includes matter-mediated contributions from the $\chi$-sector that contribute to the Yukawa coupling via anomalous dimensions. The comparison shows that trace-mode and matter-mediated effects are subleading in the regime of $g$ that obeys the weak gravity bound.}
\end{figure}

In Eq.~\eqref{eq:beta_Chi1}-\eqref{eq:etapsi}, finite $g$ results in finite fixed-point values for the corresponding couplings and anomalous dimensions. We aim at analyzing the impact of these induced interactions on $\beta_y$, i.e., ultimately on the value of $y$ at the Planck scale which is linked to $\theta_y = \left(-\frac{\beta_y}{y}\right)|_{y^\ast=0}$. Specifically,
\begin{align}
	\theta_y = 
	-\left(\eta _{\psi }+\frac{\eta _{\phi }}{2}\right)
	-g\;\mathcal{D}_y +4 \lambda_{A} \mathcal{M}_y^{\lambda y}
	\label{eq:thetaY1}
\end{align}
In the absence of gravity, \emph{all} contributions vanish at the free fixed point. Thus, the critical exponent of the Yukawa coupling at the free fixed point depends on the gravitational coupling directly, through $\mathcal{D}_1$, and indirectly: Gravity induces nontrivial anomalous dimensions $\eta_{\phi}, \eta_{\psi}$, which enter $\theta_y$, and also induces a finite four-fermion coupling $\lambda_A$, which enters $\theta_y$. The direct and the anomalous-dimension-mediated spin-2 contributions dominate the result, cf.~Fig.~\ref{fig:thetaY_contributions}.
\\
In the spin-2 approximation (where $\mathcal{D}_y = \mathcal{D}_{\lambda_A} = -5/8\mathcal{D}_{\eta_\psi}$ and $\mathcal{D}_{\eta_\phi}=0$) the suppression of matter-mediated effects can be  
demonstrated explicitly by solving Eq.~\eqref{eq:beta_lamda_A} and plugging the result back into Eq.~\eqref{eq:thetaY1}.
In a regime where gravity is not strongly dynamically suppressed by a large effective mass of the TT-propagator (i.e., $\mu_h+b-d< 2$) matter contributions are typically subleading
by a factor of $\frac{1}{16\pi^2}$ occurring in pure-matter loops. Therefore, Eq.~\eqref{eq:thetaY1} can be expanded for small $\mathcal{M}_y^{\lambda y}$ and we find
\begin{align}
	\theta_y =
	&\left(\frac{g \mathcal{D}_y-32}{8}+\frac{\left| g \mathcal{D}_y-8\right| }{2}\right)
	\notag\\&
	-\frac{16 \mathcal{M}_y^{\lambda y} \mathcal{D}_\lambda\;g^2}{\left| g \mathcal{D}_y-8\right| }
	+\mathcal{O}\left(\left(\mathcal{M}_y^{\lambda y}\right)^2\right)\;,
\end{align}
where we have also exploited that $\mathcal{M}_y^{\lambda y} = \mathcal{M}_\lambda^\lambda$.
For small $g\ll 1$ the matter-mediated $\mathcal{M}_y^{\lambda y}$-term is suppressed by the canonical dimensionality of the induced coupling $\lambda_A$. For large $g\gg 1$ the suppression relies solely on the matter-loop suppression 
\begin{align}
	\mathcal{M}_y^{\lambda y} \simeq \frac{\mathcal{D}_{y/\lambda}}{16\pi^2}\;.
\end{align}
This  explicit example is in accordance with the expectation that the gravity loop, which features several modes, dominates over the matter loops in $\beta_y$, with only one matter mode. We conclude that it is sufficient to consider only the direct contribution from the TT-mode to obtain a good understanding of the bound on the gravitational parameter space in our model.
However, the suppression of matter-mediated contributions might not hold in models with many fermions and scalars.

 \section{Observational constraints on the gravitational coupling space}
\label{sec:results}

Two observational constraints on the gravitational coupling space arise from the inclusion of matter. These constraints arise by requiring that a viable microscopic dynamics exists (cf.~Sec.~\ref{sec:weak_gravity}) and can be connected to known physics in the IR (cf.~Sec.~\ref{sec:phenoViability}).
The first constraint is due to the demand that the interacting matter fixed-point must exist, i.e., lie at real values of the couplings.
The second constraint arises when bridging the gap between the fixed-point regime and the electroweak scale: The quantum-gravity induced matter fixed-point must be phenomenologically viable in the sense that the RG flow emanating from it can lead to the observed low-energy values of the couplings.

\subsection{Weak-gravity bound}
\label{sec:weak_gravity}

In the presence of quantum gravity fluctuations, a scale-invariant fixed-point regime for matter cannot be reached in a fully asymptotically free way. Instead, all interactions which obey the symmetries of the kinetic terms for the matter fields are necessarily non-zero. 
On the other hand, those interactions with reduced global symmetries -- including all canonically marginal Standard-Model-like interactions -- can vanish in the UV. The fixed-point values for the non-zero matter couplings are functions of the gravity couplings, and are not guaranteed to be real.
Thus, the requirement of real fixed-point values for matter couplings gives rise to constraints on the microscopic gravity model.
\\
All induced interactions are canonically irrelevant, i.e., the corresponding couplings have negative canonical dimensionality.
For {induced scalar interactions}, this follows as the combination of $\mathbb{Z}_2$ and shift symmetry provides $\partial_\mu\phi\partial_\nu\phi$ as the only building block. Similarly, chiral U(1) phase rotations for fermions forbid all bilinears except for the kinetic term.
In particular, there are no matter interactions with an uneven number of matter fields. For an induced interaction with $2n$ matter fields and coupling $\bar{\lambda}_{2n}$, there is a gravity contribution $\sim g^n$ to the beta function of the dimensionless version of that coupling $\lambda_{2n} = \bar{\lambda}_{2n}k^{-d_{\bar{\lambda}_{2n}}}$, 
\be
\label{eq:general_beta}
\beta_{\lambda_{2n}} = \left(-d_{\bar{\lambda}_{2n}}+ (\mathcal{D}_{\lambda_{2n}} + \mathcal{C}_{\lambda_{2n}})g + \mathcal{M}_{\lambda_{2n}}\right)\lambda_{2n}+ \mathcal{I}_{\lambda_{2n}}\, g^n,
\ee
where $\mathcal{M}_{\lambda_{2n}}$ is a function of the other matter couplings.
\\
Crucially, the gravity-contributions $\mathcal{I}_{\lambda_{2n}}$ are independent of \emph{any} of the matter couplings. Thus, setting all matter interactions to zero does \emph{not} correspond to a fixed point of the system under the impact of gravity. The free fixed point that exists in the pure matter system is \emph{necessarily} shifted to become an interacting one: the shifted Gau\ss{}ian fixed point \cite{Eichhorn:2011pc,Eichhorn:2012va,Eichhorn:2013ug,Meibohm:2016mkp,Eichhorn:2016esv}.

We first focus on quartic interactions $\lambda_4$ (which in our truncation include $\lambda_A$ and $\chi_{1,2}$), which play a special role: These are the only matter couplings for which 
\be
\mathcal{M}_{\lambda_4}= \mathcal{M}_{\lambda_4}^{\lambda_4}\lambda_4 + \dots
\ee
 i.e., their beta functions are \emph{parabolas}. The interplay of gravity and matter fluctuations can have a disastrous effect, if the relative sign between $\mathcal{I}_{\lambda_4}$ and $\mathcal{M}_{\lambda_4}^{\lambda_4}$ is positive. Then, the shifted Gau\ss{}ian fixed point collides with a second fixed point and both move into the complex plane at $g>g_{\rm crit}$. The critical value $g_{\rm crit}$ depends on the details of the beta function. For instance, if we set all gravity couplings except for $g$ to zero, neglect anomalous dimensions and work in the spin-2 approximation,
 \bea
\label{eq:betaX1simplified}
\beta_{\chi_1} &= &4 \chi_1+ \frac{17}{84\pi^2}\chi_1^2 + \frac{80}{9}g^2 -\frac{179}{144\pi}g \chi_1 
\\
\label{eq:betaLambdaAsimplidied}
	\beta_{\lambda_A} &= & 2 \lambda_A-\frac{\lambda_A^2}{16 \pi ^2}
	+\frac{5 g^2}{4}
	+\frac{5 g \lambda_A}{2 \pi }
\eea
Both beta functions have two zeros, one of which lies at finite coupling already at $g=0$:
\bea
\chi_{1\, 1,2}^{\ast}&&=\frac{7\pi}{408} \Bigl(179\, g- 576\pi\\\nonumber
&& \pm \frac{\sqrt{-820193 g^2-1443456\pi\, g+2322432\pi^2}}{\sqrt 7} \Bigr),\\
\lambda_{A\, 1,2}^{\ast}&=& 2\pi \Bigl(8\pi +10g \pm \sqrt{105 g^2+160 \pi\, g+64 \pi^2} \Bigr).
\eea
\\
For finite $g$, both zeros lie at finite $\chi_1$ and $\lambda_A$, respectively. 
For $\lambda_A$, the shifted Gau\ss{}ian fixed point moves towards more negative values as $g$ increases, while the second fixed point moves to increasingly positive values. A fixed-point collision which would kick the two fixed points into the complex plane is averted. In contrast, the shifted Gau\ss{}ian fixed point for $\chi_1$ lies at negative values and moves towards the second fixed point as $g$ increases. At $ g_{\rm crit} \simeq 3.20$  a fixed-point collision occurs. For larger $g$, the matter fixed-point lies at complex values, i.e., it is not a viable fixed point.
This suggests that fixed-point gravity must remain sufficiently weak in order to admit a scale-invariant regime in models containing a matter sector. 
The mere existence of matter thus results in a weak-gravity bound. In the fermionic sector the negative sign in form of the fermionic loop contribution $\sim\lambda_A^2$ in Eq.~\eqref{eq:betaLambdaAsimplidied} results in a bound at imaginary $g$.

\begin{figure*}[t]
	\centering
	\includegraphics[width=0.325\linewidth]{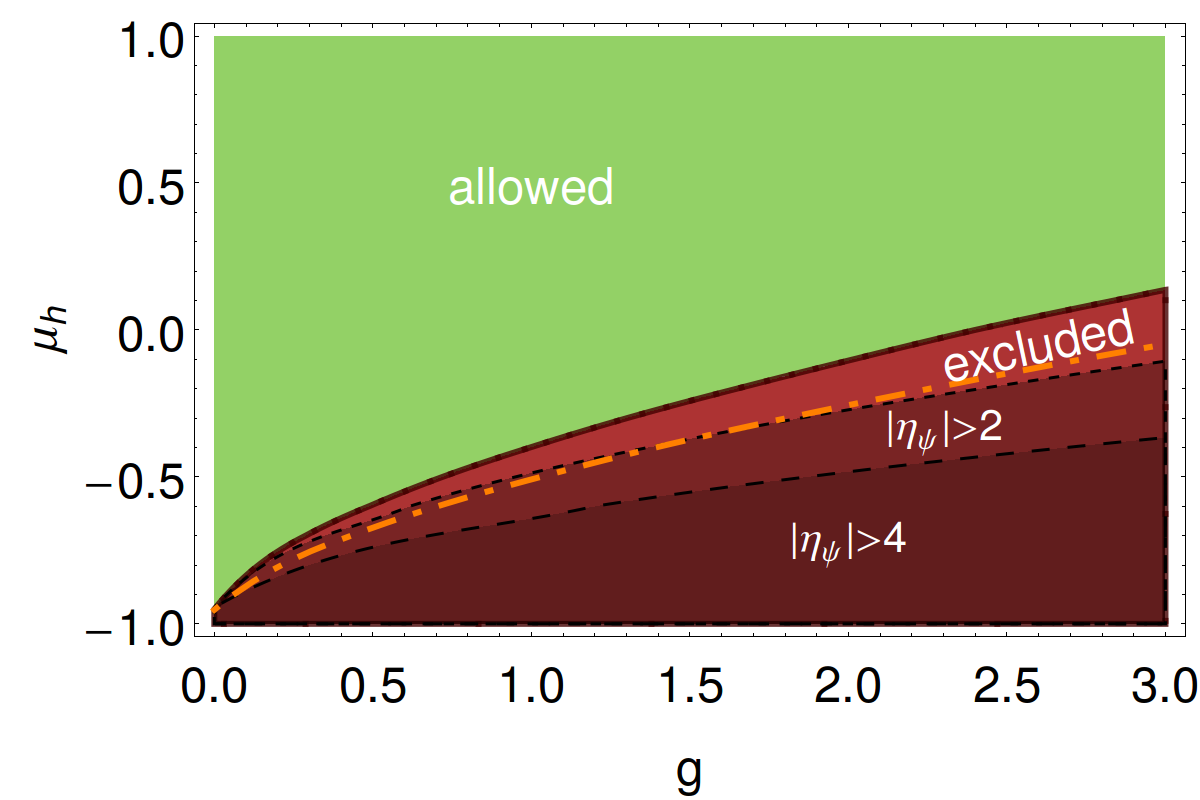}
	\includegraphics[width=0.325\linewidth]{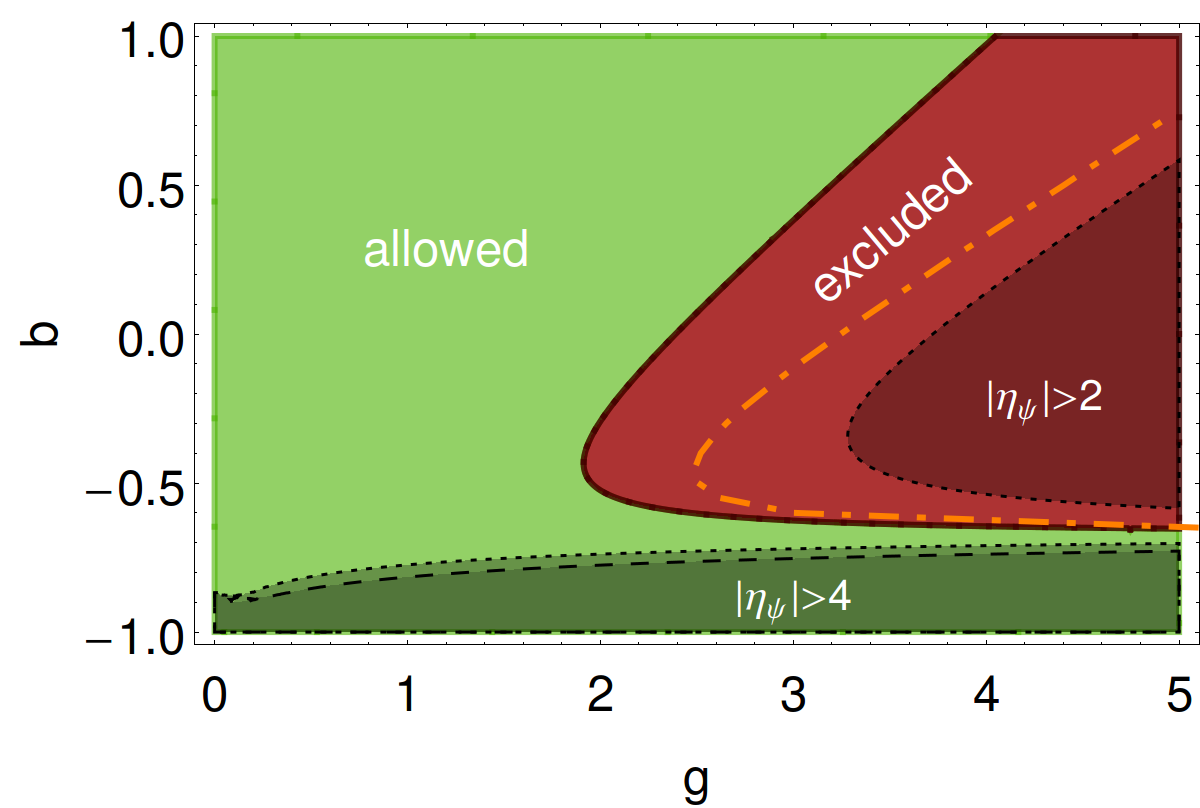}
	\includegraphics[width=0.325\linewidth]{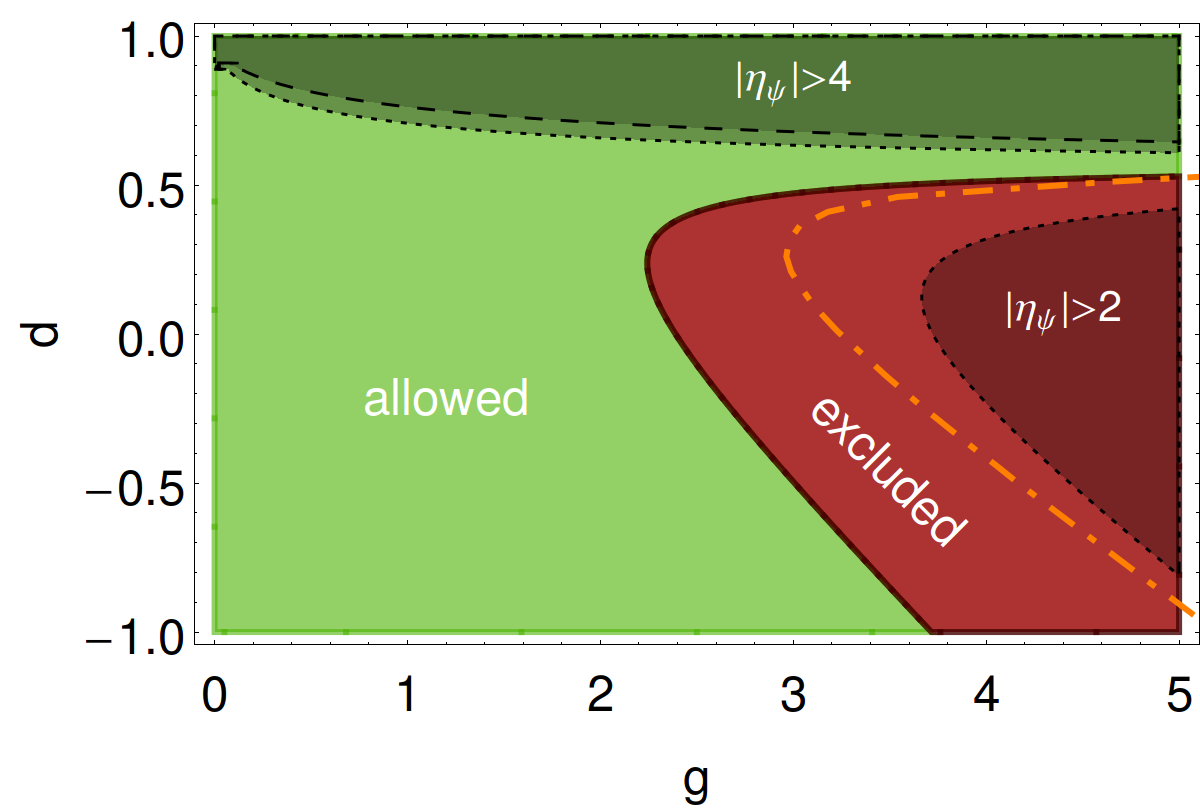}
	\caption{\label{fig:weakGravity_bound}
	 We show bounds on the gravitational parameter space $(g,\mu_h,b,d)$: In the dark red region the  spin-2-mode contributions (cf.~Sec.~\ref{sec:TTmodeApprox}) to $\beta_{\chi_{1/2}}$ are too strong to support a real fixed point in $\chi_{1/2}$. 
	 A viable fixed point exists in the light green region. We set $b=0=d$ (left panel), $\mu_h=0=d$ (central panel) and $\mu_h=0=b$ (right panel). The dot-dashed orange lines show the result when the anomalous dimensions that are shown explicitly in Eq.~\eqref{eq:etaphi} \& \eqref{eq:etapsi} are neglected. We work in the TT-mode approximation (cf.~Sec.~\ref{sec:TTmodeApprox}) and neglect anomalous dimensions in the loop diagrams. FIG.~\ref{fig:weakGravity_bound_loosendApprox} in  App.~\ref{app:TT-mode-justify} contains a comparison to the full result, highlighting the viability of our approximations. The thin dotted (dashed) lines indicate areas where $\eta_\psi<-2$ ($\eta_\psi<-4$), possibly signaling a shortcoming of our truncation. 
	}
\end{figure*}

The constraints arising from requiring real fixed-point values for the induced couplings in the gauge sector \cite{Christiansen:2017gtg} are structurally similar to those in the Yukawa sector, and reinforce our weak gravity bound. 
In addition to those couplings that we have analyzed explicitly here, all quartic couplings with higher-order momentum dependence are in danger of being shifted into the complex plane by gravity.  An analysis of the full momentum-dependent four-point vertex function of all matter fields is therefore indicated. 

Next, we shed light on the properties of induced interactions with six or more fields. These are shifted by gravity from their free fixed-point values, as in Eq.~\eqref{eq:general_beta}, but can only be shifted towards real fixed-point values, never into the complex plane. This is because in $\beta_{\lambda_{2n}}$, $\mathcal{M}_{\lambda_{2n}}$ does \emph{not} depend  on $\lambda_{2n}$ for $n>2$. Due to the one-loop structure of the flow equation (cf.~Eq.~\eqref{eq:flow}) a term $\sim\lambda_{2n}^2$ contributes to the flow of a $((2n)-4)$-coupling. Thus, only the $\beta$-function for $\lambda_4$ is quadratic in itself. Accordingly, $\beta_{\lambda_{2n}}$ is \emph{linear} in $\lambda_{2n}$ for $n>2$ and thus necessarily features a real fixed-point solution if all coefficients remain real\footnote{In principle, the linear term might cancel, but this would require a delicate balance of canonical scaling, matter fluctuations and gravity fluctuations, and hence seems not very likely. Given the one-loop structure of the flow equation (cf. Eq.~\eqref{eq:flow}) this can only couple back into its own beta function if it is combined with a quartic interaction, i.e., it cannot occur quartic or higher.  Additionally, there are linear terms from the canonical dimensionality of the coupling, anomalous scaling and a gravity contribution, schematically
\be
\beta_{\lambda_{2n}} = \left(- d_{\bar{\lambda}_{2n}}+ n\, \eta + g\, \#_1 + \lambda_4\, \#_2\right)\lambda_{2n}+...,
\ee
where the first two terms are canonical and anomalous scaling and the third and fourth correspond to a gravity- and a matter loop diagram.}. 
Therefore, fixed-point values for higher-order couplings can only become complex, when the quartic couplings become complex. Hence, no independent constraints on the gravitational parameter space arise from the requirement that the induced matter fixed-point for all couplings with six or more fields is real.

\subsubsection{Characterizing the allowed weak-gravity regime}
\label{sec:inducedBounds_thresholds}
Having motivated the existence of a weak-gravity bound, we now quantify the excluded region in gravitational coupling space in the present truncation. As a main outcome of our analysis, we find only mild changes of the bound when extending the truncation in comparison to our previous study in \cite{Eichhorn:2016esv}. 
 Away from poles in the propagating gravitational degrees of freedom (which artificially enhance the trace-mode), we only observe subleading contributions as we include the trace-mode (cf.~Fig.~\ref{fig:weakGravity_bound_loosendApprox} in App.~\ref{app:TT-mode-justify}). Since the dependence of the gauge-parameter $\beta$ is encoded in the scalar degrees of freedom this leads us to conclude that the occurrence of a weak-gravity bound does not depend on the choice of gauge.
In Eq.~\eqref{eq:general_beta}, $\mathcal{I}_{\lambda_{2n}}$ is a function of all couplings in the propagator of metric fluctuations. In Eqs.~\eqref{eq:betaX1simplified} and \eqref{eq:betaLambdaAsimplidied} we have considered the simplest case where all gravity couplings except $g$ as well as the anomalous dimensions are set to zero. In that case, the weak-gravity bound is $g^{\ast}< g_{\rm crit} \simeq 3.20$.
Reinstating the threshold corrections in $\mathcal{I}_{\chi_{1/2}}$ and $\mathcal{D}_{\chi_{1/2}}$ in Eqs.~\eqref{eq:beta_Chi1} and \eqref{eq:beta_Chi2} leads to a more involved structure of the bounds on the gravitational interaction strength. Mainly, the simple bound on $g$ is now deformed, as the additional gravitational couplings $\mu_h$, $b$ and $d$ can change the effective gravitational interaction strength at fixed $g$. Specifically, the inducing contribution $\mathcal{I}_{\chi_{1/2}}$ of spin-2-fluctuations in $\beta_{\chi_{1/2}}$ (cf.~Eq.~\eqref{eq:beta_Chi1} \& \eqref{eq:beta_Chi2}) arises from diagrams with two metric propagators (cf.~second line in Fig.~\ref{fig:flowDiagsChi}) and is therefore (at vanishing anomalous dimensions) given by
\be
\label{eq:gEff_weakGrav}
\mathcal{I}_{\chi_{1/2}} = \frac{8g^2 (10 + 15 b - 18 d)}{9(1+\mu_h+b-d)^3}= \frac{800\,g_{\rm eff}^2(1+\mu_h+b-d)}{9(10 + 15 b - 18 d)}\;.
\ee
Here, $g_\text{eff}$ is the effective gravitational interaction strength of gravitational tadpole contributions and is explicitly defined in Eq.~\eqref{eq:cond_explicit}. For $b=d=0$, large and positive $\mu_h$ leads to a suppression of metric fluctuations, thus delaying the onset of the fixed-point collision to larger $g$, cf. left panel in Fig.~\ref{fig:weakGravity_bound}. Next we analyze the dependence on $b$ and $d$: $b\rightarrow -1$ and $d\rightarrow 1$, respectively lead to an enhancement of gravity fluctuations yielding excluded regions. The additional factors of $b$ and $d$ in the numerator of Eq.~\eqref{eq:gEff_weakGrav} lead to a deviation of the excluded region as compared to the $g$-$\mu_h$-plane, cf.~central and right panel in Fig.~\ref{fig:weakGravity_bound}.
\\
Truncations for pure gravity generically feature fixed-point values in the \emph{allowed} weak-gravity regime, cf.~Tab.~\ref{tab:weakgravityyes}. The critical remaining question is to understand whether an increasing number of matter fields pushes the effective gravitational interaction strength to large values, as suggested by results in the scalar sector \cite{Dona:2013qba,Dona:2014pla,Dona:2015tnf,Meibohm:2015twa,Reichert:2017}, or whether, e.g., the impact of vector fluctuations decreases $\mathcal{I}_{\chi_{1/2}}$ sufficiently \cite{Dona:2013qba}. In the former case, bounds on the number of matter fields in asymptotically safe gravity-matter systems will arise, due to a gravity-induced ``self-destruct" mechanism: An increasing number of matter fields could push the effective gravitational interaction strength beyond the weak-gravity regime. In turn, this would lead to a fixed-point annihilation in the matter sector that would occur beyond a critical number of matter fields.

\subsection{Constraints from phenomenological viability}
\label{sec:phenoViability}

\subsubsection{Scenario A: Maximally symmetric fixed point}
The matter model that we analyze here contains only one Yukawa coupling, not several different ones like the Standard Model. However, it has in common with the Standard Model that if the Yukawa coupling is set to zero at some high scale, it remains exactly zero at all scales below. This information is enough to exclude that part of the gravitational parameter space in which the Yukawa coupling is irrelevant at its free fixed point -- barring huge effects from a curved critical hypersurface.
Specifically, if the gravity contribution to the $\beta$-function of the Yukawa coupling (cf.~Eq.~\eqref{eq:beta_y}) is positive, quantum gravity fluctuations force the Yukawa coupling to remain exactly zero until they are switched off dynamically at $k\approx M_{\rm Planck}$. This follows, as a negative scaling exponent drives a coupling towards its fixed-point value towards the IR, i.e.,
\begin{align}
	y(k) \approx y^\ast + \left(\frac{k}{k_0}\right)^{-\theta_y}\;.
\end{align}
Thus, finite quark and lepton masses (before dynamical chiral symmetry breaking in QCD) require that the Yukawa coupling is relevant, as only then finite values at the Planck scale can be accommodated while still reaching the free fixed point in the very far UV.
To be specific,
\begin{align}
\label{eq:yukawaBoundCond_thresholdStyle}
\theta_y > 0
\end{align}
needs to hold for relevance. $\theta_y$ is a  function of $(g^*,\mu_h^*,a^*,b^*,c^*,d^*)$.
Explicitly evaluating the threshold integrals (cf. Eq.~\ref{eq:thresholds_start}-\ref{eq:thresholds_end}) for spectrally adjusted cutoffs,
condition Eq.~\eqref{eq:yukawaBoundCond_thresholdStyle} for $\lambda_A=\chi_{1/2}=0$ and in the spin-2-approximation (cf. Sec.~\ref{sec:TTmodeApprox})
is given by
\begin{align}
	\label{eq:cond_explicit}
	0< -g_\text{eff} = -\frac{g (15 b-18 d+10)}{10\left(b-d+\mu _h+1\right){}^2} 
	\;.
\end{align}
For comparison, the full result is discussed in App.~\ref{app:TT-mode-justify}. In the absence of higher-order operators in the metric propagator, one would conclude that the Yukawa coupling is irrelevant as long as $g>0$. Then, it follows that $y(k\approx M_{\rm Pl})=0$, which is incompatible with a finite fermion mass at low energies.
\begin{figure}[t]
	\centering
	\includegraphics[width=\linewidth]{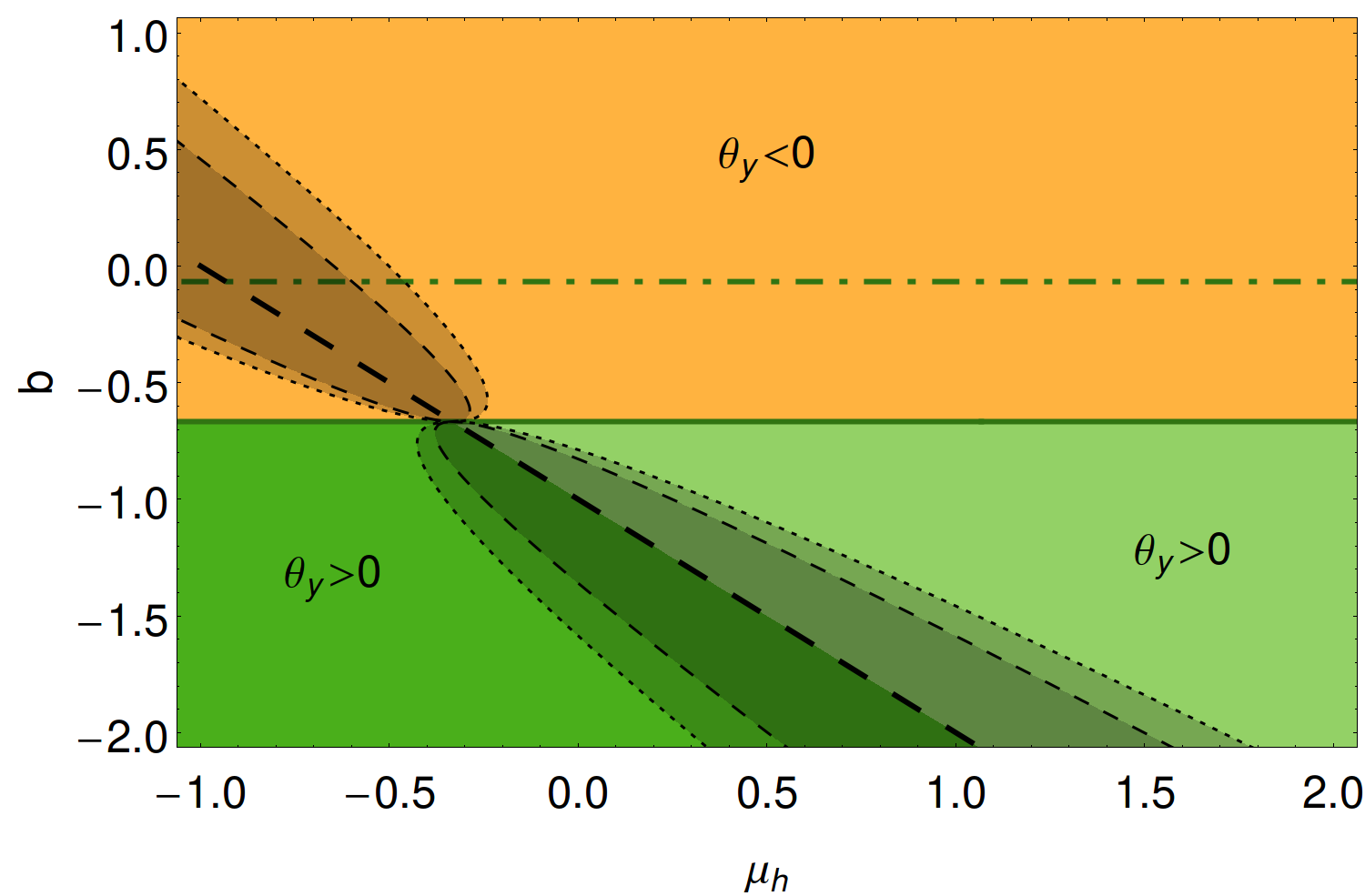}
	\caption{\label{fig:yukawaBound}
	In the green (darker) region, cond.~\eqref{eq:cond_explicit} holds (for $d=0$), and the Yukawa coupling is relevant. Thus, any low-energy value for the fermion mass can be reached on trajectories emanating from the asymptotically safe fixed point. The condition is violated in the yellow (lighter) region of the parameter space. For $d=0.5$, the boundary separating the two regions is shifted upwards (dot-dashed green line). Between the dotted (dashed) lines, the pole in the propagator in Eq.~\eqref{eq:cond_explicit} is approached, leading to $|\eta_{\psi}|>2$ (dotted line) and $|\eta_{\psi}|>4$ (dashed line), implying a possible breakdown of our truncation. For a regulator without spectral adjustment, the darker green-shaded region below the pole (thick, black, dashed line) is observationally viable, as cond.~\eqref{eq:cond_nsa} holds.
	The plot is obtained in the TT-mode approximation (cf.~Sec.~\ref{sec:TTmodeApprox})
	neglecting matter mediated effects (cf.~Sec.~\ref{sec:matterSuppression}), cf.~Fig.~\ref{fig:condition_eff} for results justifying these approximations.
	}
\end{figure}
\\
Including $b$ and $d$, the sign of the critical exponent can change, and a region in the gravitational parameter space exists which is compatible with a finite fermion mass, cf.~Fig.~\ref{fig:yukawaBound}. To highlight that this result is not an artifact of our choice of regulator, we quote the condition for a regulator that is not spectrally adjusted and where $d=0$ (for details, see App.~\ref{app:non-spectral}). Then,
\be\label{eq:cond_nsa}
0<-\frac{15 g}{16 \pi  \left(\mu _h+1\right) \left(b+\mu _h+1\right)},
\ee
which is included in Fig.~\ref{fig:yukawaBound}.
We conclude that the maximally symmetric fixed point could be observationally viable, if the fixed-point values for the gravitational couplings fall into the green region in Fig.~\ref{fig:yukawaBound}.

\subsubsection{Scenario B: Fixed point with Standard-Model symmetries}

In this subsection, we focus on a simplified truncation, where we set $\chi_{1/2}=0$,  which is sufficient for a first study of a fixed point at finite Yukawa coupling, as $\chi_{1/2}$ only couples into $\beta_y$ indirectly, i.e., through the anomalous dimensions.  
We defer the question whether the finite value of the Yukawa coupling is compatible with a fixed point for the remaining matter couplings to future work. 

The shift to relevance at the free fixed point goes hand in hand with the generation of a new, interacting fixed point for the Yukawa coupling at
\be
y^{\ast} = \sqrt{\frac{1}{\mathcal{I}_y^{y^3}}\left(-g \mathcal{I}_y + 4 \lambda_A \mathcal{M}_y^{\lambda y}-\eta_{\psi}- \frac{\eta_{\phi}}{2} \right)}.
\label{eq:NGFPy}
\ee
 For the explicit contributions $\mathcal{I}_y$, $\mathcal{I}_y^{y^3}$ and $\mathcal{M}_y^{\lambda y}$ 
 and
 the anomalous dimensions, see App.~\ref{app:thresholds}.
Where it exists, this fixed point is necessarily IR attractive, cf.~Fig.~\ref{fig:betay}, since the free fixed point is UV attractive.
\begin{figure}[!t]
\quad\quad\,\,\,\includegraphics[width=0.9\linewidth]{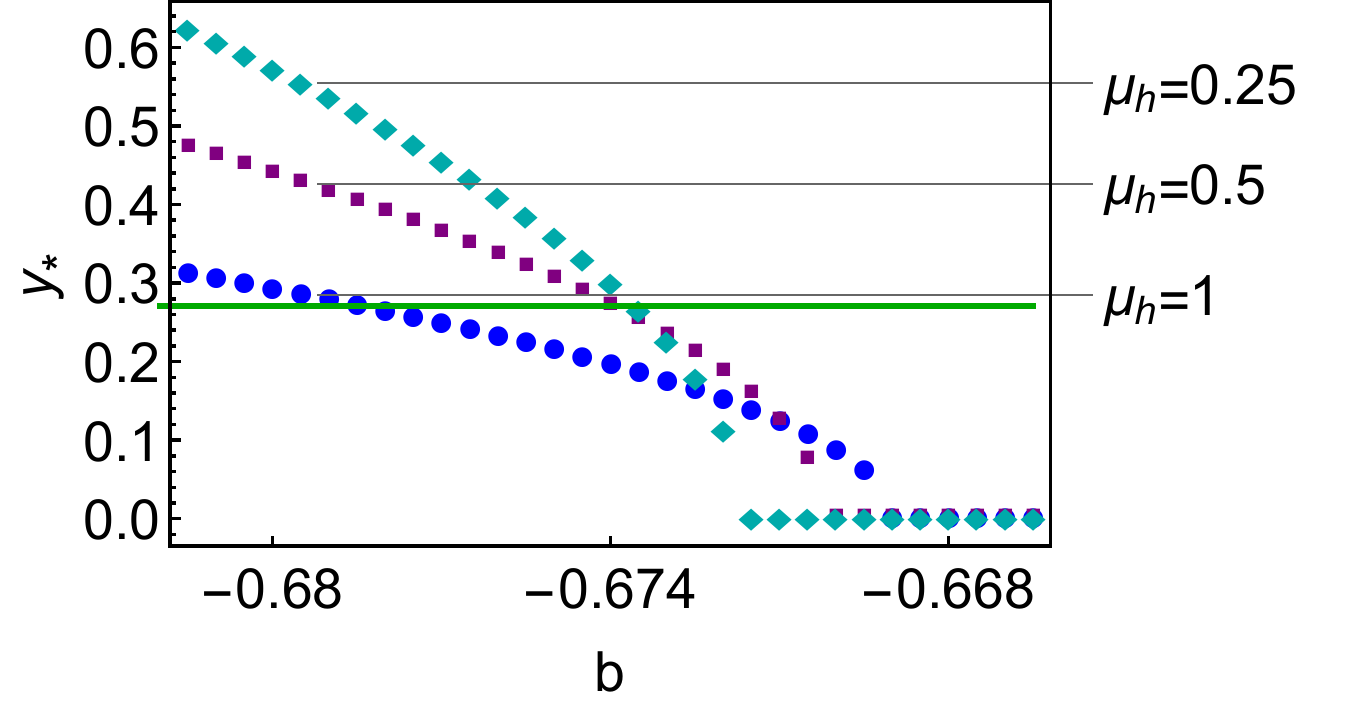}\\
\includegraphics[width=0.9\linewidth]{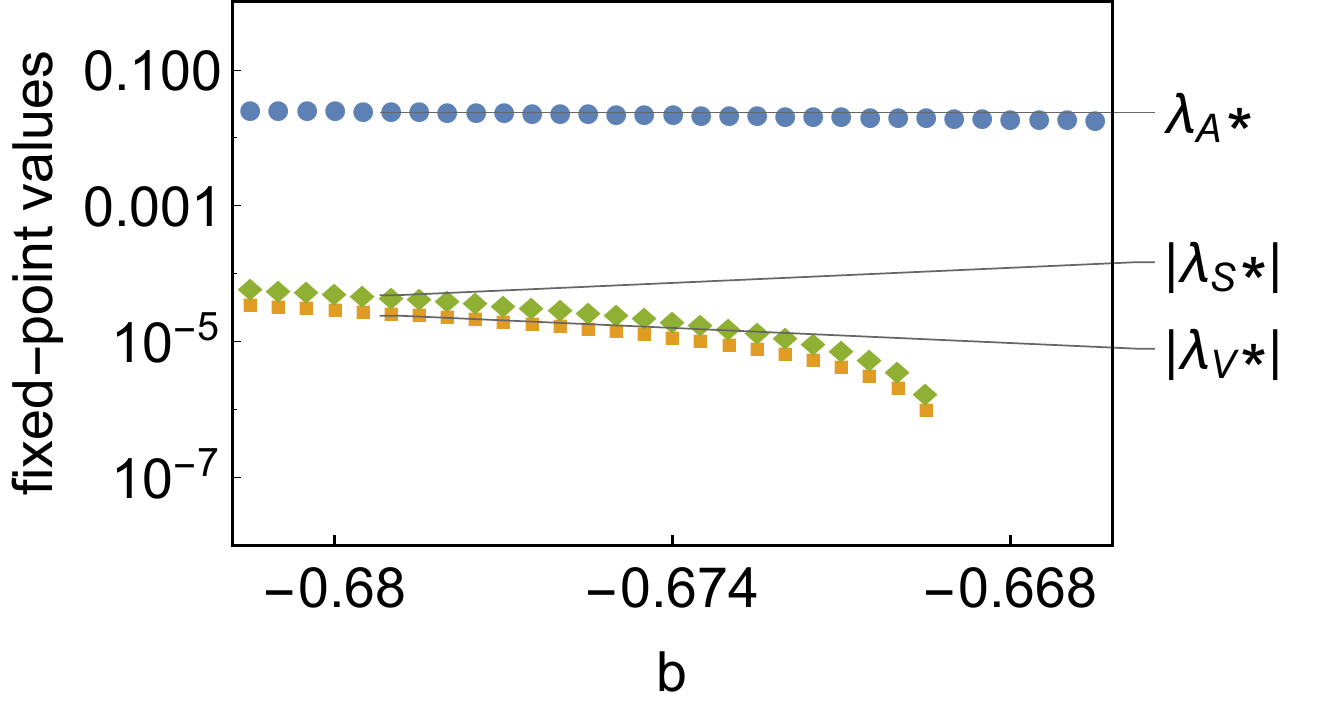}
\caption{\label{fig:NGFP} We show the fixed-point value for $y$ (upper panel) and all other matter couplings (lower panel, $\mu_h=1$) for $g=1,d=0$. 
Those couplings for which the absolute value is shown in fact have a negative sign. The green line indicates the value of the top-quark coupling in the Standard Model at the Planck scale.}
\end{figure}
\begin{figure}[!t]
\includegraphics[width=\linewidth]{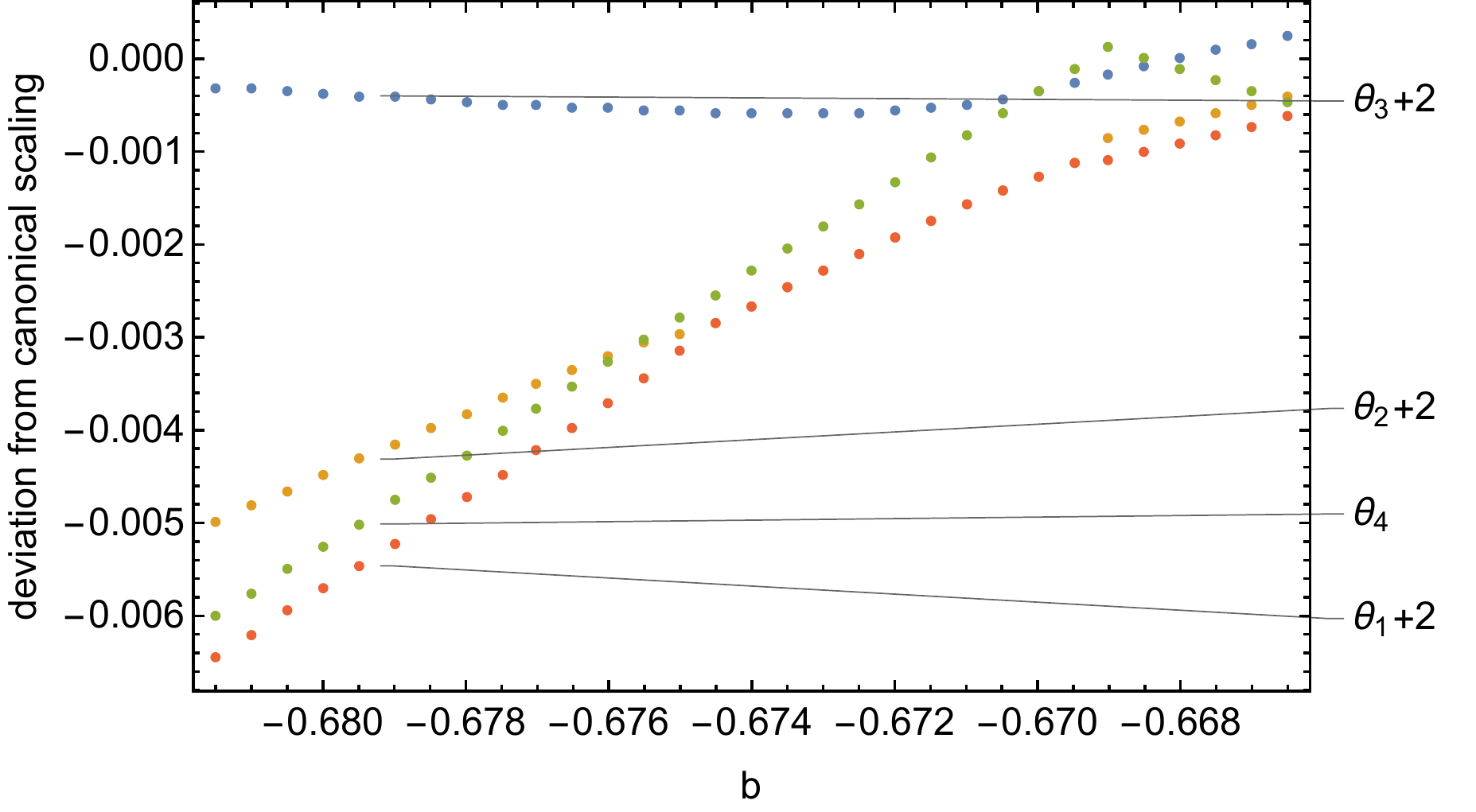}
\caption{\label{fig:thetaNGFP} We plot the deviation from the canonical dimensions for $g=1, \mu_h=1, d=0$. As the eigenvectors are rather well-aligned with the couplings $\lambda_{A, V, S}$ and $y$, we use the canonical dimensions of these couplings.}
\end{figure}

Our result highlights that the predictive power of the asymptotic-safety paradigm could be higher than that of a matter model with canonical scaling:
Starting from the interacting fixed point at $y^{\ast}$, gravity fluctuations force $y(k) \approx y^{\ast}$ until a scale is reached at which gravity is effectively switched off. The subsequent flow to the IR is the well-known pure-matter flow. Hence, $y^\ast$ translates into a fixed value of the fermion mass. Thus, there is a hypersurface in the gravitational coupling space, on which the measured value of the fermion mass in our toy model is a prediction of asymptotic safety.
As a function of $-b$, the fixed-point value in $y$ grows and quickly passes the distinguished point $y^{\ast} \approx 0.27$, cf.~Fig.~\ref{fig:NGFP}. At that point, the linear approximation to the flow in the vicinity of the fixed point  implies that the Planck-scale value of $y$ is predicted to be $y \approx 0.27$.  
Within the Standard Model, this value of the top-quark Yukawa coupling yields the measured value of the top mass in the IR, see, e.g., \cite{Degrassi:2012ry}. 
\\
As soon as the fixed point in the Yukawa coupling is interacting, all other symmetry-protected couplings, specifically $\lambda_V$ and $\lambda_S$, become nonzero, cf.~Fig.~\ref{fig:NGFP}. At this fixed point, gravity rules, i.e., corrections from induced matter interactions follow a hierarchy: $y^{\ast}$ quickly grows as a function of $-b$ and appears to reach a regime where it increases linearly. In contrast, those couplings which are already nonzero at the maximally symmetric fixed point stay nonzero, but are $\mathcal{O}(10^{-2})$. Finally, the newly generated nonzero four-fermion interactions $\lambda_S$ and $\lambda_V$ are even smaller, at $\mathcal{O}(10^{-4})$. This follows as they are not induced by gravity directly, but only through other matter-interactions, once these are generated. This
goes hand in hand with near-canonical scaling dimensions, cf.~Fig.~\ref{fig:thetaNGFP}, and a close alignment of the eigenvectors of the stability matrix with the couplings $\lambda_{A, V, S}$ and $y$.
Thus, the added complications from the matter sector only add subleading corrections to the fixed-point properties of the Yukawa coupling. We tentatively suggest that a similar hierarchy might
hold in the Standard Model, implying that the direct effects of quantum gravity will be the leading contribution to determine the scaling exponents and fixed-point values.

 \subsection{Effective field theory framework}
 \label{sec:effective}
Broadening the scope of our work, we consider other quantum-gravity models within
the gravitational parameter space from the point of view of the effective field theory (EFT) framework, \cite{Anber:2010uj,EFT}. This encompasses a large class of microscopic UV complete models.
We assume that above the scale $k_{\rm trans}$, quantum gravitational degrees of freedom are described by a fundamental model, such as, e.g., Loop Quantum Gravity, String theory, causal sets. Below $k_{\rm trans}$, effects of quantum gravity are captured in terms of quantum fluctuations of the metric, i.e., $k_{\rm trans}$ is the transition scale from the microscopic model to the effective low-energy description. In that setting, all operators which are compatible with the fundamental symmetries are present.
In line with the discussion in Sec.~\ref{sec:FPscenarios} the authors in \cite{Anber:2010uj} find that higher-order operators allowed by symmetry will be generated by quantum gravity fluctuations in the EFT framework as well. Our result might be understood as a UV completion of the EFT regime, in which the corresponding higher-order couplings cannot become asymptotically free.
The microscopic dynamics determines the effective dynamics for all effective low-energy degrees of freedom. Specifically, the microscopic model sets the initial conditions for the RG flow in the effective description, i.e., the values of all EFT couplings at $k_{\rm trans}$ are determined by the fundamental model. Typically, one would assume that $k_{\rm trans} \approx M_{\rm Planck}$, but this need not be the case. Thus, each fundamental model can be translated into a particular point, or, if it has free parameters, into a hypersurface in the space of couplings. Towards lower momentum scales, the scale-dependence is 
captured by a standard RG flow, which bridges the gap between the microscopic scale $k_{\rm trans}$ and observable scales, and connects the microscopic input-parameters at $k_{\rm trans}$ that characterize the fundamental model to low-energy parameters that can be inferred from experiments.
In this way, all quantum-gravity models where the gravitational sector reduces to General Relativity at low energies, and which do not feature any extra symmetry-violation at high $k_{\rm trans}$, can be embedded into the theory space that we analyze here\footnote{In cases with additional symmetry violation, the corresponding theory space for the low-energy model becomes larger, e.g., for the case of Lorentz-symmetry violation in the matter sector, additional couplings have to be taken into account \cite{Colladay:1998fq, Kostelecky:2003fs}. The FRG is a suitable tool to probe matter-gravity systems with such symmetry-violations \cite{Rechenberger:2012dt,Contillo:2013fua,DOdorico:2014tyh}.}. A similar point of view has already been advocated in \cite{Percacci:2010af}. 
The microscopic model predicts the values of gravity- and matter-couplings at $k_{\rm trans}$, i.e., each microscopic model gives rise to a set of predictions for the low-energy structure of the theory.
\\
There is a second scale $k_{\rm non-grav}$, at which the effect of quantum gravity fluctuations becomes strongly suppressed in  comparison to matter fluctuations, and where quantum gravity effects can be neglected. Based on dimensional arguments, one would expect $k_{\rm non-grav}\lesssim M_{\rm Planck}$. Accordingly, we assume that $k_{\rm non-grav}< k_{\rm trans}$, i.e., that there is a regime in which quantum gravity effects are important, but can be encoded within the EFT framework.

Within the functional RG setting, we analyze which regions of the gravitational parameter space are excluded, as  quantum gravity fluctuations push matter couplings away from their observed low-energy values. Specifically, we consider the perturbative regime for the Yukawa coupling $y$, where $\theta_y<0$ implies that quantum gravity pushes $y$ towards zero, as
\be
y(k) =y(k_{\rm trans})\cdot\left( \frac{k}{k_{\rm trans}}\right)^{-\theta_y},
\ee
which is clearly not compatible with large fermion masses. 
Here, we analyze a toy model, in which $y$ should not be equated with any of the Yukawa couplings of the Standard Model. Nevertheless, our analysis provides a blueprint on how to put constraints on microscopic quantum gravity models from their effect on matter.
This could lead to much stronger constraints than exist from direct tests, as, e.g., the couplings of $R^2$ and $R_{\mu \nu}R^{\mu \nu}$ are only required to be smaller than $10^{61}$ \cite{Hoyle:2004cw,Psaltis:2008bb,Atkins:2012yn}.

\section{Summary}
\label{sec:conclusions}

Here, we give a concise summary of our key findings, and 
provide
more details below. Our results are derived within a particular truncation for a toy model of the Higgs-Yukawa-sector of the Standard Model, but we argue that a significant part could carry over to extended truncations, also including additional fields.
\begin{itemize}
\item Firstly, we highlight that the seemingly daunting task of deriving the beta functions of a gravity-matter model featuring an interacting UV fixed point can be simplified considerably, as ``spin 2 rules", see subsection \ref{sec:spin2rules} below. 
\item Secondly, we clarify the interaction structure of a gravity-induced matter fixed-point and provide an ordering principle for matter interactions according to their UV fixed-point value: Scrutinizing the global symmetries of an interaction is sufficient to know whether it can feature a gravity-induced \emph{free} fixed point, see subsection \ref{sec:MSASFP} below. 
\item Thirdly, we show that the knowledge of physics at the electroweak scale could be sufficient to impose constraints on the gravitational coupling space: The mere existence of matter allows us to
derive a weak-gravity bound: If gravity fluctuations are too strong, they cannot induce a scale invariant regime for matter in the UV, and instead drive matter couplings in our truncation towards divergences. Further, recovering a finite fermion mass in the IR appears to require a negative effective gravitational interaction strength in the UV, at least within a simple Higgs-Yukawa model, see Sec. \ref{sec:boundsongrav}.
\\
Our results underline that observational constraints on quantum gravity arise without direct probes of the Planck scale: Low-energy properties of matter could be sufficient to rule out a significant part of the gravitational coupling space. Within truncations only including gravity, these parts of the gravitational coupling space do not appear special: thus, one misses these constraints if the inclusion of matter into the microscopic model is left aside.
\end{itemize}

\subsection{Spin 2 rules}\label{sec:spin2rules}
We confirm that within a large region of gravitational parameter space in our truncation spin 2 rules. This is helpful in two respects: Firstly, it allows us to derive the leading-order effects of asymptotically safe quantum gravity on matter in a simplified setting, where the number of diagrams required to gain an understanding of the fixed-point properties is reduced significantly. Secondly, it provides guidance for truncations in the gravitational sector: While truncations including more involved tensor structures are significantly more challenging computationally \cite{Benedetti:2009rx,Gies:2016con}, our results suggest that these structures dominate the quantum-gravity impact on matter.  

\subsubsection{TT mode dominance and trace mode suppression}
Our results indicate that it is sufficient to include the effect of the transverse traceless -- or spin-2- metric mode, and that, just as it should be, the scalar modes in the off-shell metric propagator are subdominant. This imbalance might be reversed in those regions of coupling space in which gravity fluctuations are suppressed at an overall level, e.g., for $\mu_h \gg 1$. Fixed points in truncations do not fall into that region, cf.~Tab.~\ref{tab:weakgravityyes} with the notable exception of a hybrid background-fluctuation calculation including the effect of all minimally coupled Standard Model fields \cite{Dona:2013qba}.

\subsubsection{Backcoupling of induced matter interactions is subleading}
At the quantitative level, we have upgraded our previous analysis in \cite{Eichhorn:2016esv} not only by higher-order gravity couplings but we now also include the effect of four-fermion couplings and anomalous dimensions. We discover that quantum-gravity induced four-fermion and two-fermion-two-scalar couplings play a subleading role for the determination of the critical exponent of the Yukawa coupling. This underpins a hope that, at least within the Higgs-Yukawa sector, a quantitative analysis of quantum-gravity effects is possible with truncations that do not include a huge number of matter interactions. Instead, we find that the leading-order quantum gravity effects are encoded in the anomalous dimensions of the matter fields, as well as the direct coupling of quantum gravity fluctuations to the Yukawa coupling. We rush to add that the suppression of all quantum fluctuations of the metric and matter besides the TT mode is highly plausible following a simple mode-counting argument at small numbers of matter fields. It is possible that this structure changes at large numbers of matter fields.
We conjecture that the suppression of the backcoupling of induced matter interactions, that we explicitly find at the current level of truncation (cf. Sec.~\ref{sec:matterSuppression} and Fig.~\ref{fig:thetaY_contributions}), can be confirmed in larger truncations.
\begin{figure}[t]
	\centering
	\includegraphics[width=\linewidth]{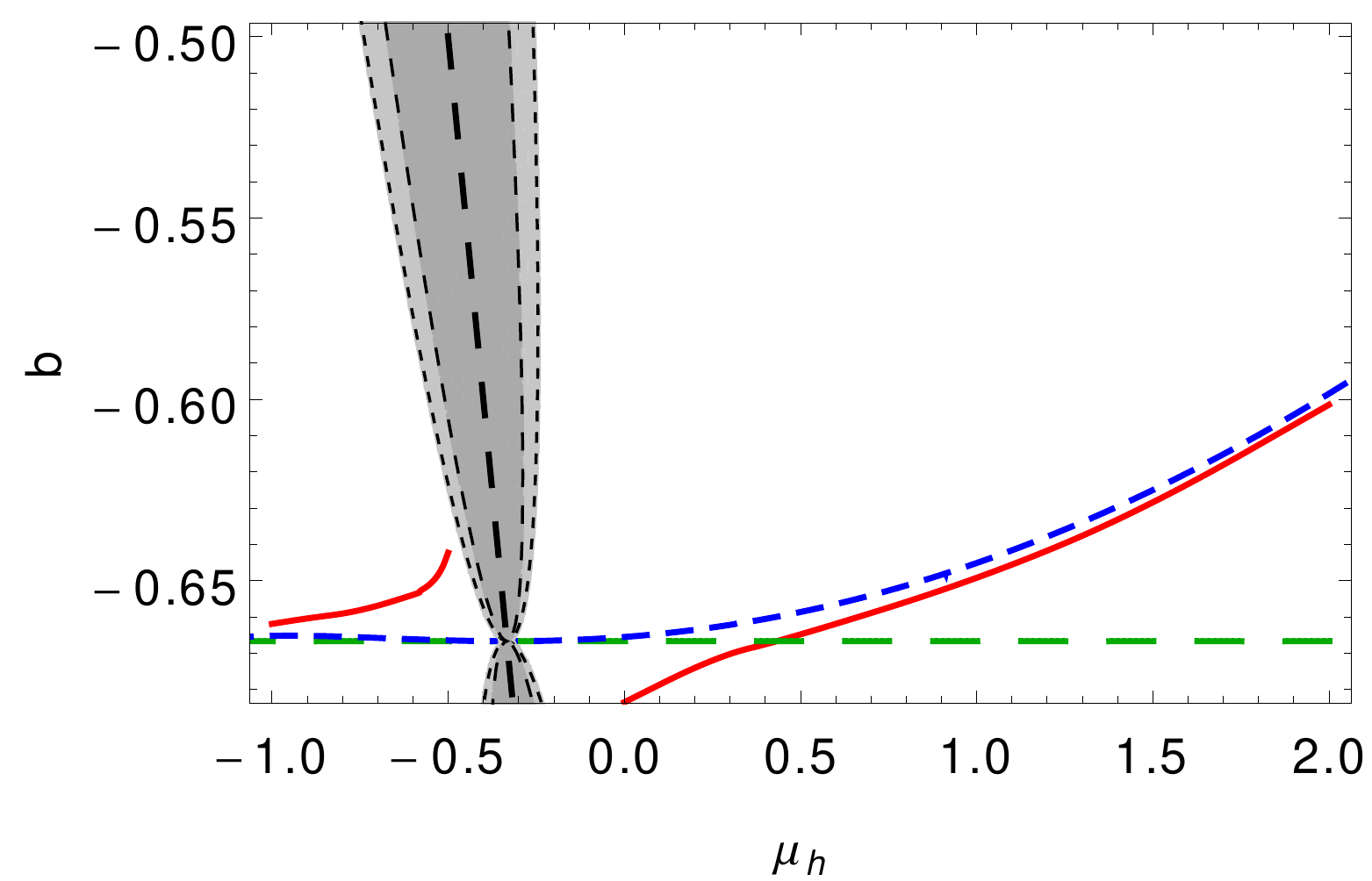}
	\caption{\label{fig:condition_eff}
	We compare the phenomenological viability bound, i.e., the relevance bound for the Yukawa coupling, for three different approximations (where the region above the respective line is excluded): the horizontal line (green wide-dashed) shows the bound in spin-two approximation (cf.~Sec.~\ref{sec:TTmodeApprox}) also neglecting all matter-mediated effects (cf.~Sec.~\ref{sec:matterSuppression}). The thick blue-dashed line shows the bound including trace-mode contributions (still neglecting matter-mediated effects). Finally, the red curve shows the full result also including matter-mediated effects from the induced couplings $\lambda_A$, $\chi_1$ and $\chi_2$. The $\chi$-sector is evaluated close to the weak-gravity bound, i.e., at $g=2$. For smaller $g$ the full result converges back to the blue-dashed line. The induced couplings become large and distort the bound whenever a pole (vertical thick dashed line) in the effective gravitational coupling strength is approached. The gray regimes around the pole indicate where anomalous dimensions become large ($|\eta_\psi|>2$ ($|\eta_\psi|>4$) along the thin (wider) dashed line). 
	}
\end{figure}
\subsection{Interaction structure of  the maximally symmetric fixed point}\label{sec:MSASFP}
Our results suggest that quantum gravity induces a fully interacting fixed point for matter. As one would expect, the finite interactions of asymptotically safe quantum gravity  percolate into the matter sector, and thus asymptotically safe gravity is incompatible with fully asymptotically free matter. We elucidate the role played by global symmetries of the kinetic terms for matter for the UV fixed point. Its interaction structure is different from that of the Standard Model: All interactions  which are compatible with the global symmetries of the kinetic terms appear to be nonzero in corresponding truncations, while those that break these symmetries explicitly vanish at the gravity-induced maximally symmetric fixed point. (The existence of the maximally symmetric fixed point does of course not preclude the existence of additional fixed points at which these global symmetries are broken explicitly and even more interactions are finite.) Specifically, the global symmetries include a shift symmetry for scalars. Thus, all induced interactions for scalars are derivative interactions, while the scalar potential, Yukawa interactions and minimal couplings of the scalar to gauge fields vanish. As the maximally symmetric fixed point does not feature any finite canonically marginal or relevant interactions, it is easy to miss the fully interacting matter fixed-point when using canonical dimensionality as a guiding principle. Such truncations neglect \emph{all} interactions which we expect to be finite at the maximally symmetric fixed point, and thus make an interacting fixed point appear free.
\begin{figure}[t]
	\centering
	\includegraphics[width=\linewidth]{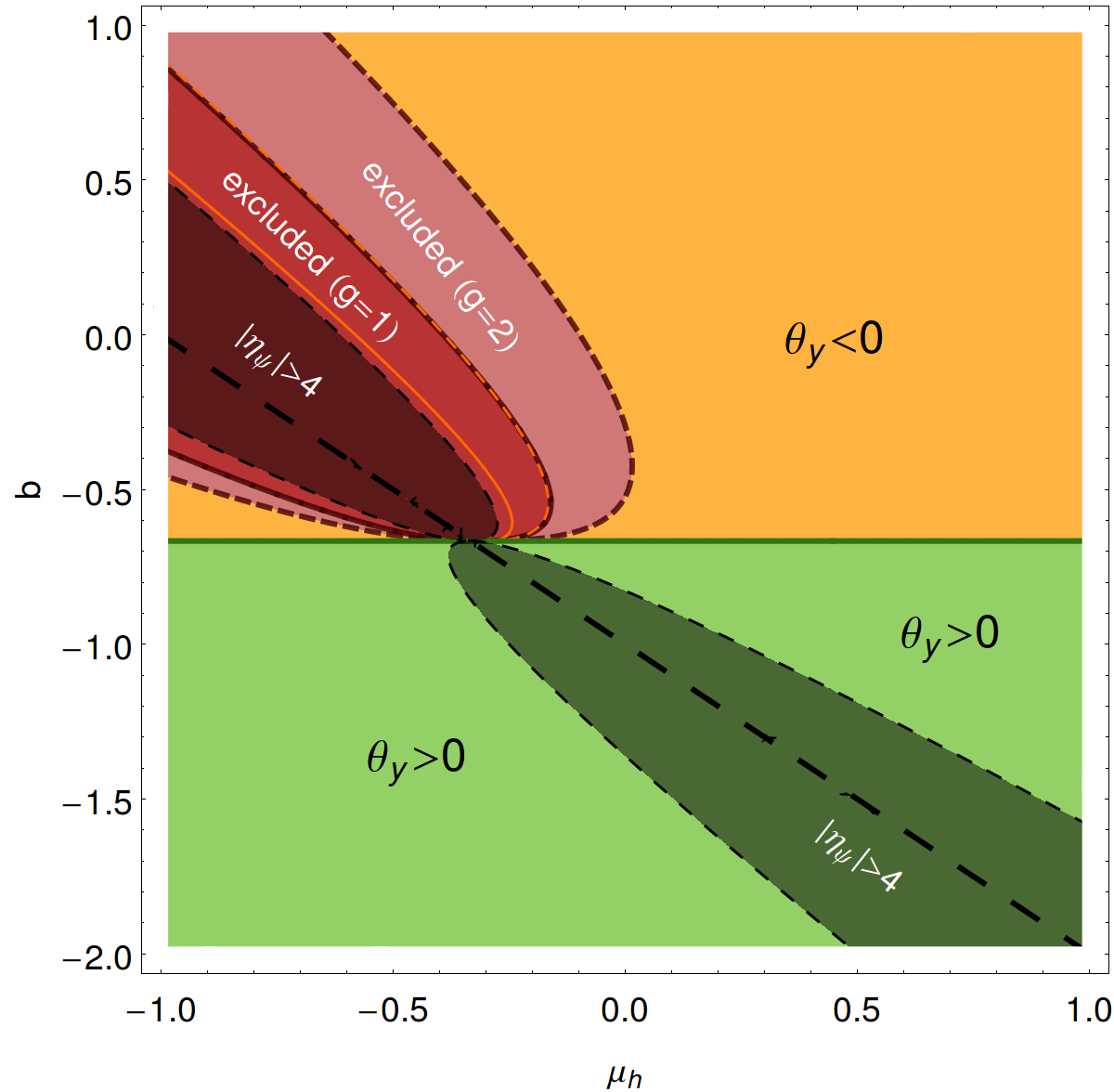}
	\caption{\label{fig:combinedBound}
	We show the	constraints on the gravitational mass parameter (related to the cosmological constant) $\mu_h$ and the higher curvature coupling $b$  in the spin-2 approximation, cf.~Sec.~\ref{sec:TTmodeApprox}). The weak-gravity bound depends on $g$: The excluded region grows with increasing $g$ (cf.~ red dashed line for $g=2$ vs.~ red continuous line for $g=1$), also	in a smaller truncation with $\eta_{\psi}=0=\eta_{\phi}$ (bright orange dashed for $g=2$ and bright orange continuous for $g=1$.) The region excluded by an irrelevant Yukawa coupling is shown in yellow. In the green region below $b \approx -2/3 $ both constraints are accommodated. Our truncation could be insufficient in the gray-shaded regions.
	}
\end{figure}

\subsection{Matter constraints on quantum gravity
}\label{sec:boundsongrav}
In the absence of direct windows into the observational quantum- gravity regime (with the notable exception of strong constraints on violations of Lorentz symmetry, see, e.g., \cite{AmelinoCamelia:1997gz,HESS:2011aa,Vasileiou:2013vra}), it becomes critical to find ways of indirectly probing quantum-gravity models. We advocate that the gap between the Planck scale and the electroweak scale can be bridged by the RG flow. We set out how to constrain physics at the Planck scale by demanding that -- starting from an asymptotically safe model of quantum gravity and matter at the Planck scale -- the experimentally observed matter properties are recovered at the electroweak scale. 

\subsubsection{Existence of a UV completion -- Weak gravity bound}
\begin{table*}[!t]
\begin{tabular}{l|c|c|c|c|c|c|c|c|c}
ref. and truncation & weak gravity & $\theta_y$ & $g^{\ast}$ & $\mu_h^{\ast} = -2 \Lambda^{\ast}$ & $b^{\ast}$& $d^{\ast}$ & $a^{\ast}$& $c^{\ast}$&$\eta_{\psi}$\\
\hline\hline
{\bf pure gravity in Einstein-Hilbert truncation}\\\hline
\cite{Gies:2015tca}
for all $\alpha=\beta=0$& yes & -0.82 & 0.88 &-0.36 & -- & --& -- & --&-1.16\\\hline\hline
{\bf pure gravity in extended truncations}\\ 
{ bimetric truncations indicated by (B)}\\\hline
\cite{Becker:2014qya} (B) & yes & -0.80 & 0.7 & -0.42 & -- & -- & -- & --&-1.14\\\hline
\cite{Benedetti:2009rx} &yes & 1.71 & 1.97 &-0.42 & -0.99&--&0.42& -- &5.04\\ \hline
\cite{Christiansen:2016sjn} (B) & yes & -0.60 & 0.43 & -0.34 & 0.39 & --& -0.18 & -- &-1.04\\\hline
\cite{Denz:2016qks} (B) & yes & -0.69 & 0.83 & -0.45& 0 & -- & -- & --&-1.00\\ \hline\hline
{\bf Einstein-Hilbert gravity with matter} \\\hline
\cite{Dona:2013qba} with Standard Model matter & yes & 0.004 & 1.74 & 4.7 &--  & -- & --&--& 0.01\\
\hline
\cite{Meibohm:2015twa} (B) for $N_S=1=N_D$ & yes & -0.83 & 0.55 & -0.58 &--  & -- & --&--& -1.25 \\\hline
{\bf gravity with matter in extended truncations}\\ \hline
\cite{Hamada:2017rvn}higher-derivative gravity with $N_S=1=N_D$& yes & 0.85 & 0.95 & 0.46 & -1 &0 & 1.19 & 0 & 1.41\\ \hline \hline
\end{tabular}
\caption{\label{tab:weakgravityyes}Here we list results from the literature and attempt to infer whether asymptotically safe gravity passes both bounds. The equality $\mu_h^{\ast} = -2 \Lambda^{\ast}$ holds within the single-metric approximation. For all truncations, the value of $\eta_{\psi}$ that we provide, is the value that we obtain from our calculation. It differs from values quoted in \cite{Dona:2013qba} and \cite{Meibohm:2015twa} due to a different choice of gauge.
\\
The answer on the weak-gravity bound appears to be an unequivocal yes.
\\
The decision on the (ir)relevance appears less clear; bimetric truncations seem to have a tendency to $\theta_y<0$, although the value is numerically small; background calculations including higher-order interactions or matter fluctuations seem to favor $\theta_y>0$ instead. Overall, we conclude that truncations have not yet converged sufficiently to make a definite statement.
Observe that no fixed-point results on $d$ exist yet. According to our analysis, this coupling could have a pivotal effect.
}
\end{table*}

If quantum fluctuations of gravity are too strong, they can neither be balanced by matter fluctuations nor  by canonical scaling in our truncation. Thus, if the effective gravitational coupling exceeds a critical strength, no UV completion can be induced in the matter sector. 
Hence, we expect a viable asymptotically safe model of quantum gravity \emph{must} satisfy the weak-gravity bound, unless matter degrees of freedom only emerge in the IR. The effective gravitational coupling parameterizes the strength of quantum-gravity fluctuations, i.e., it is a combination of the Newton coupling with additional couplings that enter the propagator of metric fluctuations. Gravity becomes strong if either the Newton coupling exceeds a critical strength, and/or the denominator of the propagator of metric fluctuations is driven towards zero. (Within a bimetric setting that is parameterized in terms of the background and fluctuation field the couplings $a,b,c,d$ should be understood as those parameterizing the higher-order momentum dependence of the fluctuation propagator, i.e., those are part of the inessential couplings.) Within our truncation, this excludes a sizable part of the gravitational coupling space in which gravity is not weak enough, cf.~red regions in Fig.~\ref{fig:combinedBound}. It is reassuring that the weak-gravity bound is satisfied by the fixed-point values in all truncations listed in Tab.~\ref{tab:weakgravityyes}.
Notably, the weak-gravity bound is \emph{not} visible in truncations including only the canonical Standard Model couplings 
due to symmetry. Instead, it arises from those couplings that share the global symmetries of the kinetic terms. Although those are higher-order couplings according to a canonical counting, their inclusion in matter-gravity truncations appears to be mandatory for studies that explore whether asymptotic safety is a viable paradigm for a UV completion of gravity-matter models.
\subsubsection{Phenomenological viability -- negative effective gravitational coupling}
The maximally symmetric UV fixed point lies in a symmetry-enhanced hypersurface of the theory space on which all canonical interactions of the Standard Model vanish. That hypersurface is IR attractive or repulsive under the RG flow. The former case provides predictions for the values of the Standard-Model couplings at the Planck scale. In our analysis, we focus on a Yukawa coupling in a simple Higgs-Yukawa model and demand a finite fermion mass, generated from spontaneous symmetry breaking in the IR. This provides a constraint on the gravitational parameter space, as the fixed point at vanishing Yukawa coupling must be UV attractive. 

Further, we explore whether fixed points with explicitly broken global symmetries, such as, e.g.,
shift-symmetry can exist, and find that within our truncation the answer is indeed ``yes" in a subspace of the gravitational parameter space. For the Yukawa coupling in our simple Higgs-Yukawa model, such a fixed point at finite $y^{\ast}$ is necessarily IR attractive. This provides us with a lower-dimensional hypersurface in the gravitational parameter space on which the Planck-scale value of the Yukawa coupling is predicted from the underlying fixed-point theory, leading to a prediction of the fermion mass in the IR.

Within a simple truncation containing only an Einstein-Hilbert term from which one can derive a metric propagator and a Yukawa term, the critical exponent of the Yukawa coupling at the free fixed point is negative cf.~Tab.~\ref{tab:weakgravityyes}, unless the fixed-point value for the Newton coupling is negative. This results in a \emph{prediction} of the fermion mass in the IR: Quantum gravity fluctuations force the Yukawa coupling to remain zero all the way to the Planck scale. Following a Standard-Model like RG flow from the Planck scale down to the electroweak scale then results in a vanishing fermion mass. The extended analysis that we have performed here confirms the result from the above simple truncation: A finite fermion mass can be accommodated in the region of gravitational coupling space with a negative effective Newton coupling. The overall result on $\theta_y$ from various truncations of the RG flow appears ``too close to call", cf.~Tab.~\ref{tab:weakgravityyes}, underscoring a need for extended truncations. Due to spin-2 dominance, this result cannot be altered by the contributions from induced matter interactions such as four-fermion couplings or fermion-scalar couplings. While this might not be an observationally desirable outcome, it underscores how coupling quantum gravity to matter moves the model a decisive step closer towards becoming falsifiable. This result will most likely carry over to more extended models of the Yukawa sector featuring several Yukawa couplings. This hinges on the fact that the gravity contribution to the Yukawa coupling is in fact ``blind" to the existence of several generations and different Yukawa couplings and and appears to work against asymptotic freedom as long as the spin-2-mode dominates.

\subsection{Outlook}
Here, we attempt to provide a roadmap towards tightening observational constraints on asymptotically safe gravity and matter along the lines that we have mapped out in this paper. In particular, we highlight which features of our analysis we expect to hold in more realistic matter models, and what are possible missing contributions beyond our truncation.

\subsubsection{Extended tests of the weak-gravity bound}
We argue that pure-matter $n$-point interactions with $n \geq6$ do not impose any further bounds on the space of gravity couplings, since we expect that gravity can only shift their fixed-point values \emph{along} the real axis. The backreaction of induced 6-point interactions into the flow of the 4-point couplings might nevertheless affect our estimate of the excluded ``strong-gravity" region quantitatively. Further, an independent weak-gravity bound could also originate at higher orders in the derivative expansion. We suggest to explore the full momentum dependence of induced matter 4-point interactions to quantitatively pinpoint the weak-gravity bound.
\\
We expect the weak-gravity bound to hold in models with  $N$ matter fields, where the combined matter-gravity dynamics could yield  an upper bound on $N$: If the fixed-point value for the effective gravity coupling scales with $N$, as expected from \cite{Dona:2013qba, Meibohm:2015twa,Dona:2015tnf}, the gravity-induced destruction of quantum scale-invariance in the matter system sets in at a critical $N_\text{crit}$.

\subsubsection{Impact of non-minimal kinetic terms}
We expect non-minimal ``kinetic" couplings, e.g., $\bar{\psi}\gamma_{\mu}\nabla_{\nu}\psi R^{\mu \nu}$ and $\partial_{\mu}\phi \partial_{\nu}\phi R^{\mu\nu}$ to potentially play a significant role for both constraints that we have explored. These couplings share the symmetries of the respective kinetic terms, and will accordingly feature a shifted Gau\ss{}ian fixed point, potentially resulting in an independent weak-gravity bound. Further, such couplings will impact the anomalous dimensions through graviton loops. In accordance with our result in the spin-2-dominated region of gravitational coupling space, these contributions might compete with the gravity contributions to the anomalous dimension that we have included in our analysis.

\subsubsection{Finite fermion mass from a relevant Yukawa coupling}
The results on fixed-point values for the metric propagator have not converged sufficiently to conclude whether the Yukawa coupling is relevant, cf.~Tab.~\ref{tab:weakgravityyes}. 
Additional options for the Yukawa coupling to become relevant could involve additional degrees of freedom: As a first possibility, the gravity couplings might move into the allowed region of parameter space as a function of the number of matter fields. Within a more realistic model of the Higgs-Yukawa-sector of the Standard Model, several different Yukawa couplings should be distinguished. Then, the region in gravitational parameter space which is compatible with realistic values of those couplings in the IR, could change.

In the Standard Model the gauge couplings yield a contribution to the running of the Yukawa coupling
\be
\beta_{y}\Big|_{\rm gauge} = \frac{y}{16 \pi^2}\left(-8 g_s^2 - \frac{9}{4}g_2^2-\frac{17}{12}g_1^2 \right),
\ee
where $g_s$ is the strong coupling, $g_2$ the SU(2) coupling and $g_1$ the Abelian hypercharge coupling. As gravity contributions tend to make gauge theories asymptotically free \cite{Daum:2009dn,Folkerts:2011jz}, even in the Abelian case \cite{Harst:2011zx,Christiansen:2017gtg}, the combination of gravity and charged matter could induce an interacting fixed point for the Abelian gauge coupling. Thus, this yields a contribution to the critical exponent of the Yukawa coupling, which is of the \emph{opposite} sign as the gravity contribution. We will estimate whether this contribution could be sufficient to render the Yukawa coupling relevant already in a truncation where $\mu_h=b=d=0$. For simplicity, we will use the beta function for the QED coupling $e$ at one loop. Including the leading-order quantum-gravity contribution it reads
\be
\beta_{e}= -\frac{5\, g}{18\pi}e + N_f\, \frac{e^3}{12\pi^2},
\ee
where $N_f$ is the number of charged fermions. The central piece of information for the following is that the potential interacting fixed point is IR attractive. Thus, the Planck-scale value of that coupling -- which is known within the Standard Model -- must be equal to the fixed-point value. Hence we can infer directly that the $\theta_y|_{\rm gauge} \approx 0.003$. Moreover,
solving the equation $e^{\ast} \approx 0.6$ (where we have for simplicity used the Planck-scale value of the hypercharge coupling) provides us with a value for the Newton coupling, $g^{\ast}\approx 0.7$ (for $N_f=21$). In turn, this yields $\theta_y|_{\rm grav} \approx -0.2$ in the TT approximation. Accordingly we conclude that an interacting fixed point for the Abelian gauge coupling provides a contribution towards a relevant Yukawa coupling, but our simple estimate suggests that  the contribution is insufficient to balance the quantum-gravity contribution. Including higher-order couplings in the metric propagator might reduce the quantum-gravity contribution and tip the balance in favor of the gauge contributions.

We point out that at leading order the direct quantum-gravity contribution to the running Yukawa coupling and quartic Higgs self interaction is exactly the same -- since both arise from a vertex derived from $\sqrt{g}$. Hence, a setting in which the Yukawa coupling becomes relevant while the Higgs quartic coupling stays irrelevant and is thus automatically held close to zero all the way down to the Planck scale as in \cite{Shaposhnikov:2009pv,Bezrukov:2012sa} relies on more involved higher-order effects, e.g., mediated by anomalous dimensions or induced couplings. The same is true for a solution of the gauge hierarchy problem in asymptotic safety \cite{Wetterich:2016uxm}, which would require relatively strong gravitational effects driven by the same vertex.
We plan to come back to this point in the future. 
\\
If the preservation of global symmetry within asymptotically safe gravity that has been observed in the structure of all beta functions up to now turns out to be an artifact of an approximation after all, then the maximally symmetric fixed point might be shifted and might in fact feature a finite fixed-point value for the Yukawa coupling.

If on the other hand, the Yukawa couplings indeed remain irrelevant in the fully coupled Standard Model plus gravity at a free fixed point, this would be an indication that new physics which does not contain a fundamental scalar for the Higgs might be required. For such a
composite Higgs, with a sufficiently high compositeness scale to evade experimental constraints,
the question whether asymptotically safe quantum gravity is compatible with the observed masses of the fermions in the Standard Model, would presumably take an entirely different form.

Our analysis highlights that for an understanding of the quantum-gravity effects on the Higgs-Yukawa sector it is critical to understand the fixed-point values for the gravitational couplings $R_{\mu\nu}\Box^n R^{\mu\nu}$, while operators including the Ricci scalar, like $R \Box^n R$, $n\geq0$, are subleading. This should provide an incentive to extend gravity truncations in these directions in the future.


\noindent\emph{Acknowledgements:} 
\\
We acknowledge helpful discussions with N.~Christiansen, S.~Lippoldt, J.~M.~Pawlowski and  C.~Wetterich, as well as with M.~Yamada on \cite{Hamada:2017rvn}.
We thank the African Institute for Mathematical Sciences in Tanzania for hospitality during part of this project. This work was supported by the DFG under the Emmy-Noether program, grant  no.~EI-1037-1. A.~E.~is also supported by an Emmy-Noether visiting fellowship at the Perimeter Institute for Theoretical Physics. Research at Perimeter Institute is supported by the Government of Canada through Industry Canada and by the Province of Ontario through the Ministry of Economic Development \& Innovation. A.~Held is also supported by a scholarship of the Studienstiftung des deutschen Volkes.

\appendix
\section{Threshold-integrals}
\label{app:thresholds}
Here we present explicit expressions for the contributions to the $\beta$-functions in Eq.~\eqref{eq:beta_Chi1}-\eqref{eq:beta_lamda_S} in terms of threshold integrals. Traces 
of the tensor structure of flow diagrams resulting from appropriate projections of the flow equation Eq.~\eqref{eq:flow} are already executed, such that only a regularized loop-integral over the propagating modes remains. Traces have been checked using the FormTracer package \cite{Cyrol:2016zqb}. Our notation for the threshold integrals essentially counts the number of propagators of a particular field, i.e., $I = I[n_\text{TT},n_\text{Tr},n_\psi,n_\phi;n_p]$. $n_\text{TT}+n_\text{Tr}+n_\psi+n_\phi$ equals the number of vertices of the corresponding diagram, see Eq.~\ref{eq:general_cutoff_integral} below. The pure matter contributions (independent of the gravitational couplings) are given by
\begin{align}
	\label{eq:thresholds_start}
	\mathcal{M}_y^{\lambda y} &= \mathcal{M}_\lambda^\lambda = 
	I[0,0,2,0;0] =- \frac{\left(5-\eta _{\psi}\right)}{80 \pi ^2}
	\\[2ex]
	\mathcal{M}_y^{y^3}&= \mathcal{M}_\lambda^{\lambda y^2} = 
	-I[0,0,2,1;0]=
	\frac{\left(5-\eta _{\psi }\right)}{80 \pi ^2}
   +\frac{\left(6-\eta _{\phi }\right)}{96 \pi ^2}
   \\
   \mathcal{M}_{\chi_1}^{\chi^2} &=
   -\frac{\left(121 \chi_1^2+64 \chi_1 \chi_2+4 \chi_2^2\right)}{24}  I[0,0,(1),1;3]
   \notag\\
   &
   -\frac{\left(\chi_1-2 \chi_2\right) \left(5 \chi_1+2 \chi_2\right)}{4} I[0,0,1,(1);3]
   \notag\\
   &=
   \frac{\left(121 \chi_1^2+64 \chi_1 \chi_2+4 \chi_2^2\right) \left(9-\eta _{\phi }\right)}{6048 \pi ^2}
   \notag\\
   &
   +\frac{\left(\chi_1-2 \chi_2\right) \left(5
   \chi_1+2 \chi_2\right) \left(8-\eta _{\psi }\right)}{1792 \pi ^2}
   \;.
	\\[2ex]
   \mathcal{M}_{\chi_2}^{\chi^2} &=
   \frac{\left(59 \chi_1^2-52 \chi_1 \chi_2-76 \chi_2^2\right)}{48} I[0,0,(1),1;3]
   \notag\\
   &
   +\frac{\left(\chi_1-2 \chi_2\right) \left(\chi_1+\chi_2\right)}{2} I[0,0,1,(1);3]
   \notag\\
   &=
   -\frac{\left(59 \chi_1^2-52 \chi_1 \chi_2-76
   \chi_2^2\right) \left(9-\eta _{\phi }\right)}{12096 \pi ^2}
   \notag\\
   &
   -\frac{\left(\chi_1-2 \chi_2\right) \left(\chi_1+\chi_2\right) \left(8-\eta _{\psi }\right)}{896 \pi ^2}
   \\[2ex]
   \mathcal{M}_{\eta_\phi}^{y^2} &= \frac{y^2 \left(4-\eta _{\psi }\right)}{16 \pi ^2}
	\\[2ex]
	\mathcal{M}_{\eta_\phi}^\chi &=
	4(\chi_1 + 4\chi_2) \,I[0,0,1,0;1] = 
	-\frac{(\chi_1 + 4\chi_2) \left(6-\eta _{\psi }\right)}{60 \pi ^2}
	\\[2ex]
   	\mathcal{M}_{\eta_\psi}^{y^2} &= y^2 \frac{\left(5-\eta _{\phi }\right)}{80 \pi ^2}
   \;.
	\\[2ex]
	\mathcal{M}_{\eta_\psi}^\chi &=
	-\frac{\chi_1 + 4\chi_2}{2}\,I[0,0,0,1;2] = 
	\frac{(\chi_1+ 4\chi_2) \left(8-\eta _{\phi }\right)}{384 \pi ^2}
   \;.
\end{align}
The gravity contributions depend on the exact form of the gravity propagator. In terms of threshold functions of the propagating degrees of freedom -- i.e., the TT- and trace-mode -- the induced contributions read
\begin{align}
	\mathcal{I}_{\chi_1} &=	
		-\frac{5}{72}I[2,0,0,0;0]
		-\frac{3}{32768}I[0,2,(1),1;3]
		\notag\\&
		+\frac{9}{16384}I[0,(2),1,(1);3]
   	\\[2ex]
   	\mathcal{I}_{\chi_2} &=
		\frac{5}{288}I[2,0,0,0;0]
		+\frac{3}{16384}I[0,2,0,1;2]
		\\&\notag
		-\frac{57}{65536}I[0,2,(1),1;3]
		-\frac{9}{16384}I[0,(2),1,(1);3]
   	\\[2ex]
   \mathcal{I}_\lambda &=
   -\frac{15}{1024}I[2,0,0,0;2]
   \;.
\end{align}
Further gravity contributions linear in the corresponding matter coupling itself appear as a shift in the canonical scaling dimension. The corresponding contributions in terms of threshold integrals are given by
\begin{align}
   	\mathcal{D}_{\chi_1} &=		
		\frac{5\left(31 \chi_1+16 \chi_2\right)}{288} I[1,0,0,0;0]
		\notag\\&
		+\frac{5\left(\chi_1-4 \chi_2\right)}{72} I[1,0,1,0;1]
		\notag\\&
		-\frac{\chi_1}{128}I[0,1,0,0;0]
		-\frac{27\chi_1}{512}I[0,1,(1),0;1]
		\notag\\&
		-\frac{3\chi_1}{128}I[0,(1),1,0;1]
		-\frac{\chi_1}{64}I[0,1,0,1;2]
		\notag\\&
		+\frac{27\chi_1}{512}I[0,1,(2),0;2]
		+\frac{\chi_1}{192}I[0,1,0,2;4]
		\notag\\&
		+\frac{8\chi_1+\chi_2}{128}I[0,1,(1),1;3]
		\notag\\&
		-\frac{3(\chi_1+\chi_2)}{64}I[0,(1),1,(1);3]
   \;,
   \\[2ex]
   	\mathcal{D}_{\chi_2} &=		
		\frac{5\left(8 \chi_1+23 \chi_2\right)}{288} I[1,0,0,0;0]
		\notag\\&
		-\frac{5 \left(\chi_1-4 \chi_2\right)}{288} I[1,0,1,0;1]
		\notag\\&
		-\frac{\chi_2}{128}I[0,1,0,0;0]
		-\frac{3(\chi_1+13\chi_2)}{512}I[0,1,(1),0;1]
		\notag\\&
		-\frac{3(\chi_1+8\chi_2)}{512}I[0,(1),1,0;1]
		-\frac{\chi_2}{64}I[0,1,0,1;2]
		\notag\\&
		+\frac{9(\chi_1+7\chi_2)}{512}I[0,1,(2),0;2]
		\notag\\&
		+\frac{9(\chi_1+4\chi_2)}{512}I[0,(1),2,0;2]
		\notag\\&
		+\frac{\chi_1+6\chi_2}{384}I[0,1,0,2;4]
		\notag\\&
		+\frac{13\chi_1+38\chi_2}{512}I[0,1,(1),1;3]
		\notag\\&
		+\frac{3(\chi_1+4\chi_2)}{256}I[0,(1),1,(1);3]
\end{align}
\begin{align}
	\label{eq:D_y}
	\mathcal{D}_y &= 
		-\frac{5}{4} I[1,0,0,0;0]
   		+ \frac{1}{16} I[0,1,0,0;0] \nonumber
   	\\&	- \frac{3}{16}  I[0,1,1,0;1]
   		+ \left(\frac{3}{16}\right)^2 I[0,1,2,0;2]
   	\\[2ex]
   	\mathcal{D}_\lambda &=
   		-\frac{5}{4}I[1,0,0,0;0]
   		+ \frac{1}{16}I[0,1,0,0;0] \nonumber
   	\\&	-\frac{3}{8}I[0,1,1,0;1]
   		+ \frac{27}{128}I[0,1,2,0;2]
   	\\[2ex]
   	\label{eq:D_etaPhi}
	\mathcal{D}_{\eta_\phi} &= 	
		\frac{1}{64}I[0,1,0,1;2]
	\\[2ex]
	\label{eq:D_etaPsi}
   \mathcal{D}_{\eta_\psi} &=
   		\frac{25}{32}I[1,0,0,0;0]
   		-\frac{3}{128} I[0,1,0,0;0]
   		\notag\\&
   		+\frac{117}{1024}I[0,1,(1),0;1]
   		+\frac{9}{128}I[0,(1),1,0;1]
   		\;.
\end{align}
Finally, mixed contributions from gravity and other matter couplings are given in terms of threshold integrals as
\begin{align}
\mathcal{C}_\lambda^{gy^2} &=
	\frac{1}{4}I[0,1,0,1;0]
	+\frac{3}{8}I[0,1,1,1;1]\nonumber
	\\& +\frac{9}{64}I[0,1,2,1;2]
	\label{eq:thresholds_end}
	\;.
\end{align}
In all above equations
$I[n_\text{TT},n_\text{Tr},n_\psi,n_\phi;n_p]$ represents the scale derivative of the corresponding threshold function of a loop integral for $n_\text{TT}$ traceless-transverse metric, $n_\text{Tr}$ trace metric, $n_\psi$ fermionic and $n_\phi$ scalar propagating modes. Further $n_p$ gives the power of additional momenta from the vertices involved. Finally, the round brackets indicate that the scale derivative $\partial_t$ does not act on the corresponding propagator; or put in a diagrammatic language that the diagram with the corresponding regulator insertion is excluded.
Therefore,
\begin{align}
	\label{eq:general_cutoff_integral}
	&I[n_\text{TT},n_\text{Tr},n_\psi,n_\phi;n_p] = \tilde{\partial}_t \int\frac{d^4 p}{(2\pi)^4}\left(p^2\right)^{(n_p+n_\psi)/2}\Bigg[
	\notag\\&\quad
	\frac{1}{\left[Z_\psi p^2\left(1+r_{k,f}\left(\frac{p^2}{k^2}\right)\right)\right]^{n_\psi}}
	\times\frac{1}{\left[Z_\phi p^2\left(1+r_{k,b}\left(\frac{p^2}{k^2}\right)\right)\right]^{n_\phi}}
	\notag\\&\quad
	\frac{1}{\left[\Gamma_\text{k,TT}^{(2)}\left(1+r_{k,b}\left(\frac{p^2}{k^2}\right)\right)\right]^{n_\text{TT}}}
	\times\frac{1}{\left[\Gamma_\text{k,Tr}^{(2)}\left(1+r_{k,b}\left(\frac{p^2}{k^2}\right)\right)\right]^{n_\text{Tr}}}
	\Bigg]\;.
\end{align}
In this form it is straightforward to associate diagrams with the different contributions: For $I[n_\text{TT},n_\text{Tr},n_\psi,n_\phi]$, the diagram has $n_\text{TT}+n_\text{Tr}+n_\psi+n_\phi$ vertices, and the power of different propagators of course simply corresponds to the number $n_i$.
\\
Using spectrally and RG-adjusted cutoffs (cf.~Eq.~\eqref{eq:cutoff-spectrally}) to evaluate the threshold integrals (cf.~Eq.~\eqref{eq:general_cutoff_integral}) a general threshold integral can be evaluated to read
\begin{widetext}
\begin{align}
	\label{eq:general_thresholdIntegral}
	&I[n_\text{TT},n_\text{Tr},n_\psi,n_\phi;n_p] =
	\frac{2^{8 n_\text{Tr}+5 n_{\text{TT}}-3} \pi ^{n_\text{Tr}+n_{\text{TT}}-2}}{n_p+4} 
	\left(\frac{1}{b-d+\mu _h+1}\right)^{n_{\text{TT}}} 
	\left(\frac{1}{18 a+6 b-18 c-6 d-2 \mu _h-3}\right)^{n_\text{Tr}}
	\\&\notag\quad
	\Bigg[
	-\frac{2 n_{\text{TT}} \left(2 b \left(n_p+6\right)
   \left(n_p+10\right)-\left(n_p+8\right) \left(3 d \left(n_p+6\right)-n_p-10\right)\right) \left(-\eta _h+n_p+6\right)}{\left(n_p+6\right) \left(n_p+8\right) \left(n_p+10\right) \left(b-d+\mu
   _h+1\right)}
   	\\&\notag\quad\;\;
   	-\frac{6 n_\text{Tr}\left(-\eta _h+n_p+6\right) \left(12 a \left(n_p+6\right) \left(n_p+10\right)+4 b \left(n_p+6\right) \left(n_p+10\right)-\left(n_p+8\right) \left(18 c \left(n_p+6\right)+6 d
   \left(n_p+6\right)+n_p+10\right)\right)}{\left(n_p+6\right) \left(n_p+8\right) \left(n_p+10\right) \left(18 a+6 b-18 c-6 d-2 \mu _h-3\right)}
   \\&\notag\quad\quad
   -\frac{n_{\psi } \left(-\eta _{\psi }+n_p+5\right)}{n_p+5}-\frac{2 n_{\phi } \left(-\eta _{\phi }+n_p+6\right)}{n_p+6}\Bigg]\;,
\end{align}
\end{widetext}
where contributions in the square brackets correspond to contributions where the scale-derivative acts on the TT, the trace, the fermionic and the scalar propagating mode respectively. As explained above, the prefactors that arise from a contraction of the tensor structure corresponding to a diagram, can differ for diagrams within the same class, where the regulator is inserted on different internal legs. These are the contributions which feature round brackets around some of the arguments of the $I$.
Thus, where round brackets are denoted in Eq.~\eqref{eq:thresholds_start}-\eqref{eq:thresholds_end} the corresponding term within the square brackets in the above equation must be neglected.
\\
Note that in the spin-2 approximation ($n_\text{Tr}=0$) the threshold integrals only depend on the gravitational parameters $(g,\mu_h,b,d)$ but not on $(a,c)$ which only occur in the trace-mode.
Substituting Eq.~\eqref{eq:thresholds_start}-\eqref{eq:thresholds_end} into the $\beta$-functions in Eq.~\eqref{eq:beta_Chi1}-Eq.~\eqref{eq:beta_lamda_S} and evaluating the threshold integrals according to Eq.~\eqref{eq:general_thresholdIntegral} gives the set of gravitationally modified $\beta$-functions used to obtain all the results in this paper. We loosen the spin-2-approximation in App.~\ref{app:TT-mode-justify}. In App.~\ref{app:non-spectral} we evaluate the cutoff integrals for a non-spectral choice of regulator.

\section{Regulator dependence}
\label{app:non-spectral}
\begin{figure}[t]
	\centering
	\includegraphics[width=\linewidth]{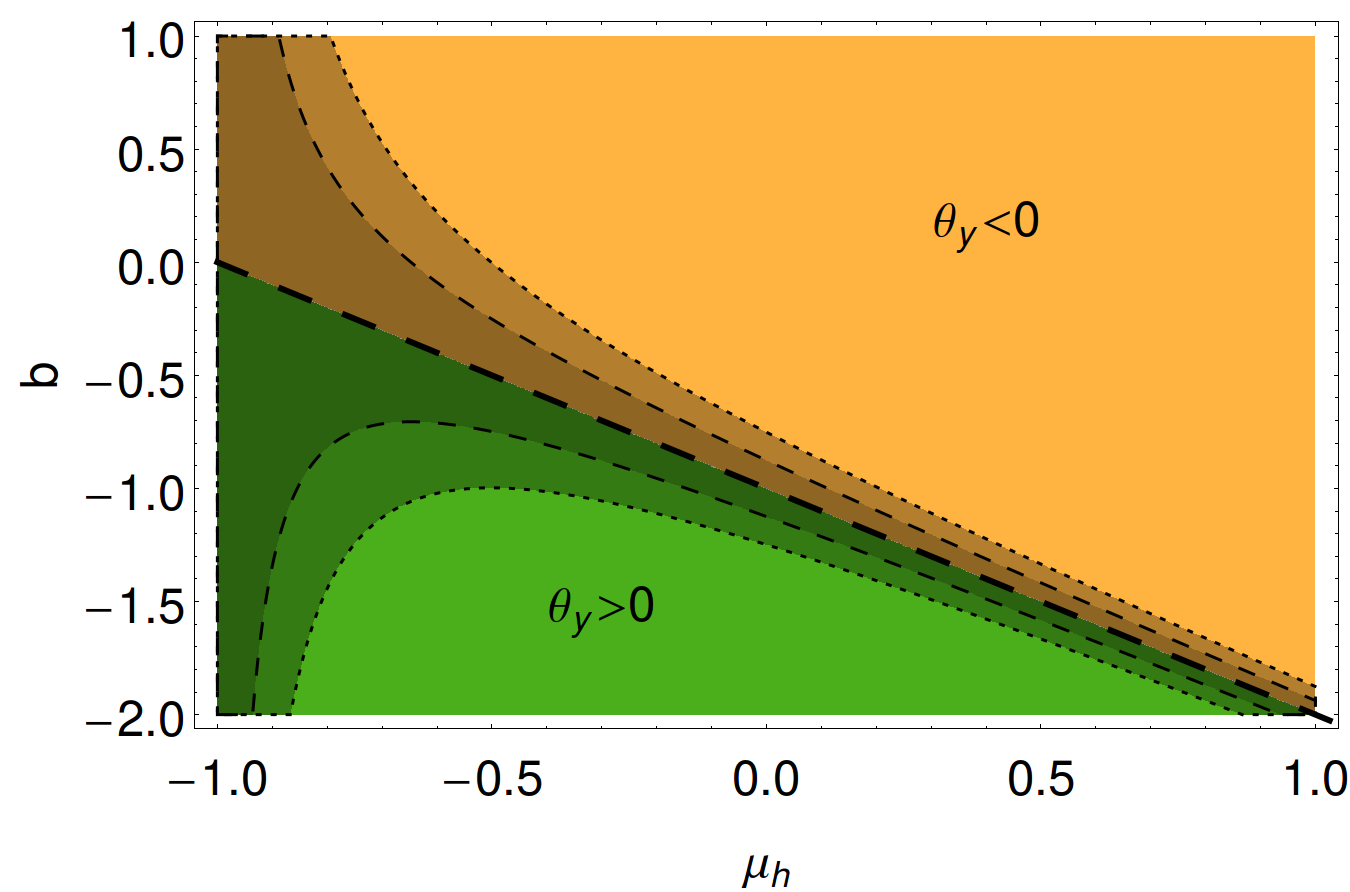}
	\caption{\label{fig:yukawaBoundNonSpectrally}
	We show the phenomenological viability bound on gravitational coupling space consisting of $\mu_h, b, d$ in the TT approximation for a non-spectral cutoff choice.
	Here we set $d=0$. In the yellow (brighter shaded) region $\theta_y<0$ at the maximally symmetric fixed point. In the green (darker shaded) region $\theta_y>0$ at the maximally symmetric fixed point. We show contour lines of potentially increasing truncation instability quantified by $|\eta_\psi|>2$ (dotted thin line) and $|\eta_\psi|>4$ (dashed thin line) as one moves towards the pole in the gravitational TT-mode (thick dashed line). The sign change in $\theta_y$ is affected by a pole-crossing in the TT-mode propagator.
	}
\end{figure}
Here, we evaluate the threshold-integrals relevant for the phenomenological viability bound (cf.~Sec.~\ref{sec:phenoViability}) in the TT-approximation (cf.~Sec.~\ref{sec:TTmodeApprox}) neglecting matter-mediated contributions (cf.~Sec.~\ref{sec:matterSuppression}), i.e., Eq.~\eqref{eq:D_y}, \eqref{eq:D_etaPhi} and \eqref{eq:D_etaPsi} and derive the bound that requires weakly interacting gravity for a non-spectrally adjusted cutoff. Explicitly, the cutoff takes only $p^2$ as the argument, instead of $\Gamma_k^{(2)}(p^2)$, as in the spectrally adjusted case. As the shape function, we use the Litim-shape function. The Yukawa bound in the TT approximation Eq.~\eqref{eq:yukawaBoundCond_thresholdStyle} leads to the condition
\begin{align}
\label{eq:yukawaBoundCond_nonSpectral}
-\frac{15 g}{16 \pi  \left(\mu _h+1\right) \left(b+\mu _h+1\right)}>0\;,
\end{align}
where we have set $d=0$. We plot this condition in Fig.~\ref{fig:yukawaBoundNonSpectrally}.
It is obvious that in this cutoff scheme the sign change required to render $y$ relevant occurs along the poles of condition Eq.~\ref{eq:yukawaBoundCond_nonSpectral}. 
\\
It is reassuring that the two different cutoff schemes exhibit the same limits and both admit
a relevant Yukawa-coupling in a largely overlapping region of parameter space.
In other words, the exclusion of $b>0$ is not due to our choice of regulator. 

\section{Beyond the TT-mode approximation}
\label{app:TT-mode-justify}

\begin{figure*}[t]
	\centering
	\includegraphics[width=0.49\linewidth]{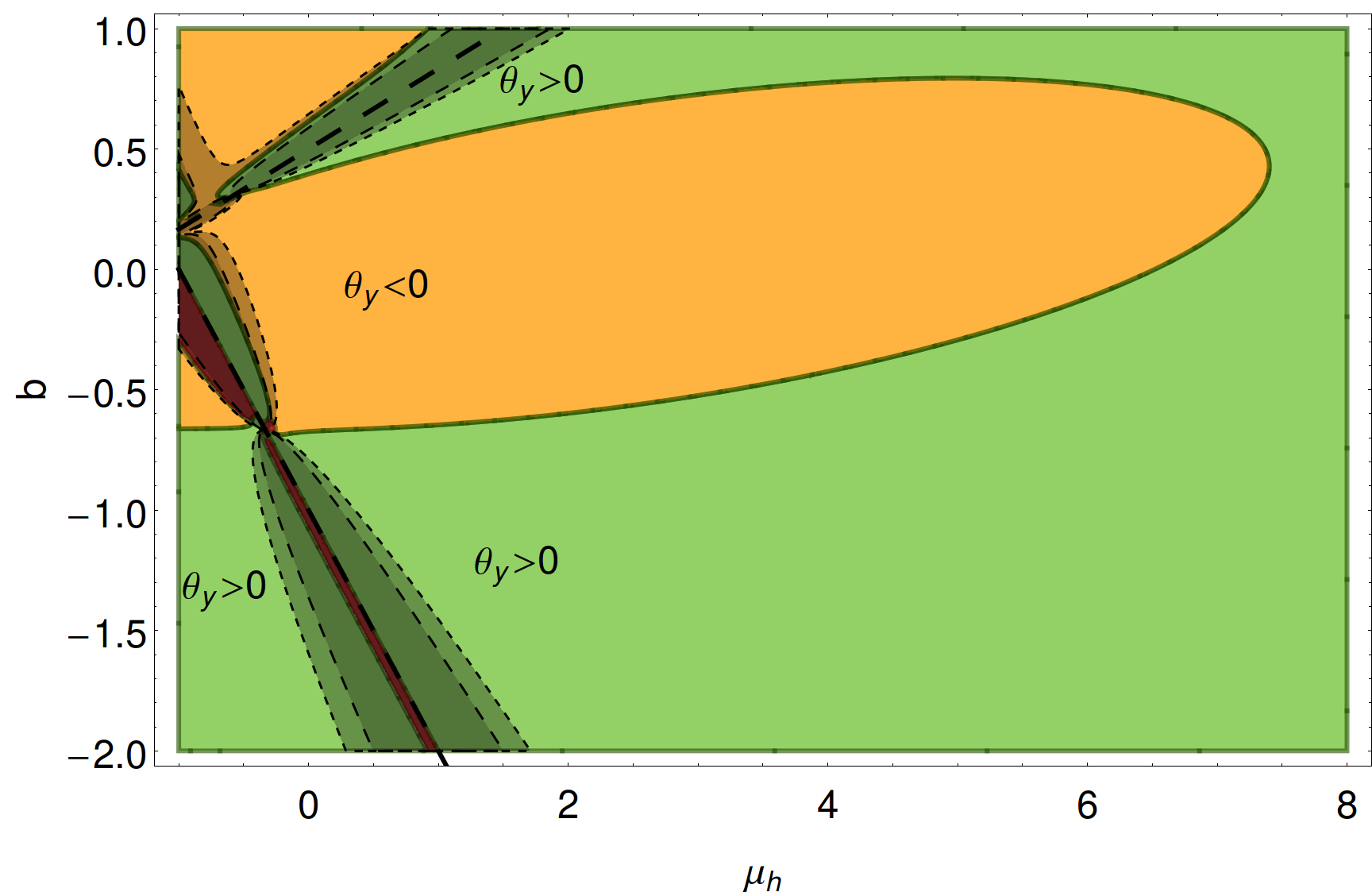}
	\includegraphics[width=0.49\linewidth]{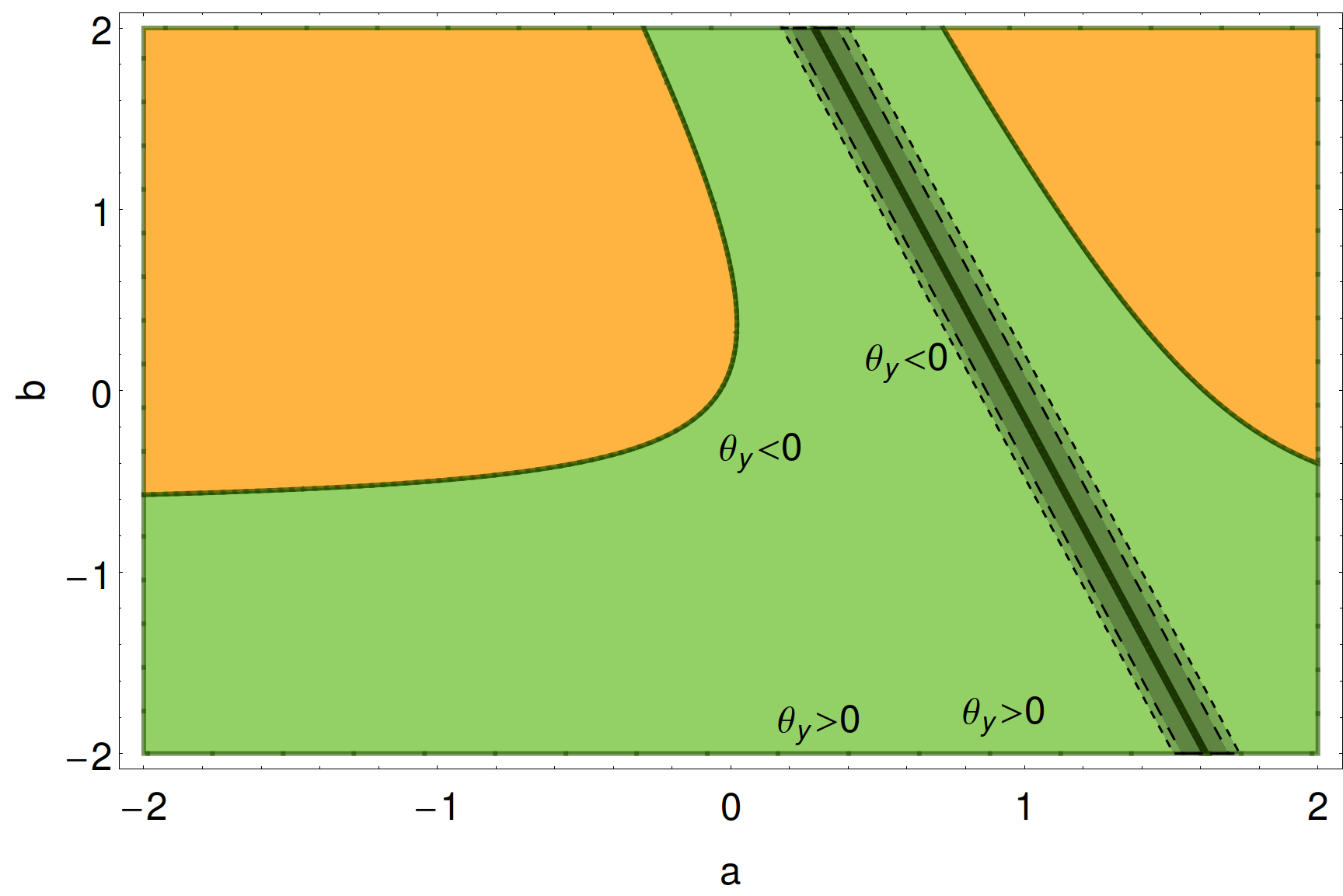}
	\caption{\label{fig:condition_eff}
	We plot the (ir-)relevance of the Yukawa coupling in (yellow) green regimes at the maximally-symmetric asymptotically safe fixed-point lifting the following approximations (employed in FIG.~\ref{fig:yukawaBound}): we include the trace-mode contributions; we solve the full set of equations for the anomalous dimensions (including loop-terms); we include effects from $\beta_{\lambda_A}$. The left-hand plot shows a $\mu_h$-$b$-slice through the gravitational coupling space ($a=c=d=0$, $g=1$).	Along the poles of the spin-2- and trace-mode (thick dashed black lines) the gray-shaded regions ($|\eta_\psi| > 2$ (dotted line) and $|\eta_\psi| > 4$ (dashed line) indicate a possible breakdown of our truncation. In the red (dark) region within this area (and beyond the TT-mode pole) $\lambda_A$ develops a weak-gravity bound (because $\mathcal{I}_{\chi_{1/2}}$ in Eq.~\ref{eq:gEff_weakGrav} switches sign). The right-hand panel shows a similar $a$-$b$-slice through gravitational parameter-space ($\mu_h=c=d=0$, $g=1$). We observe that the spin-2-mode contribution and therefore the influence of $b,d$ and $\mu_h$ dominate also in the extended analysis, as long as $\mu_h$ does not become large. The dependence on $a$ and $c$ becomes relevant only in a regime (very thin slice in right-hand panel) where parameters are tuned to artificially enhance the trace-mode.	}
\end{figure*}

\begin{figure*}[!t]
	\centering
	\includegraphics[width=0.28\linewidth]{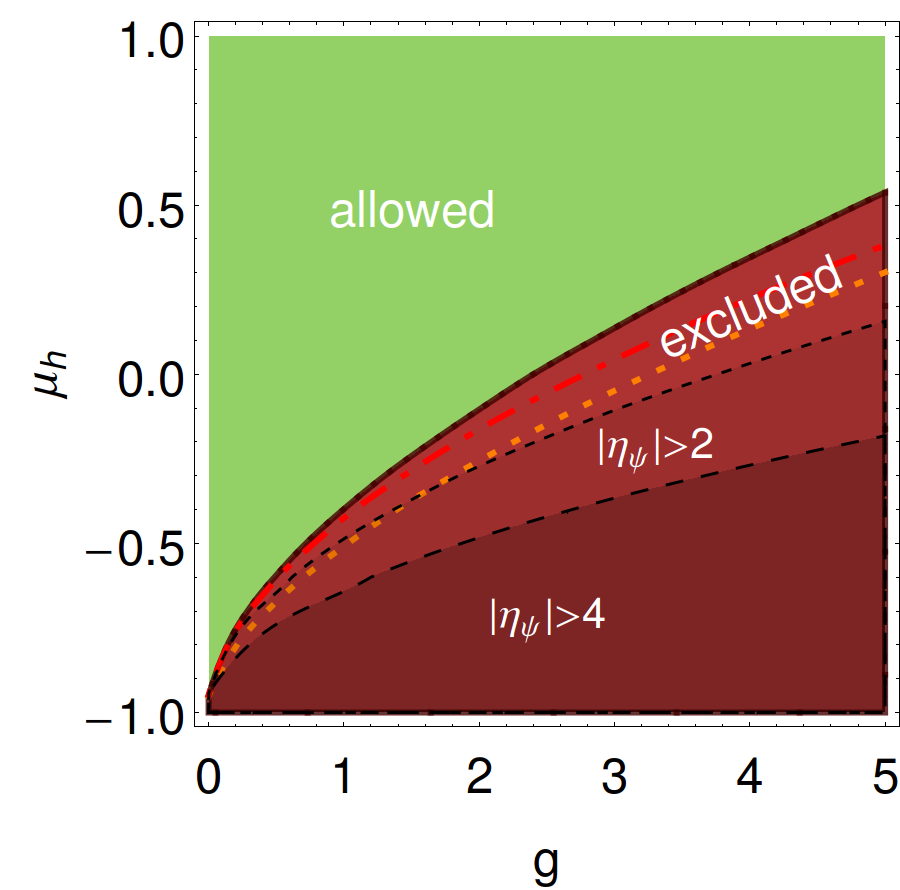}
	\includegraphics[width=0.28\linewidth]{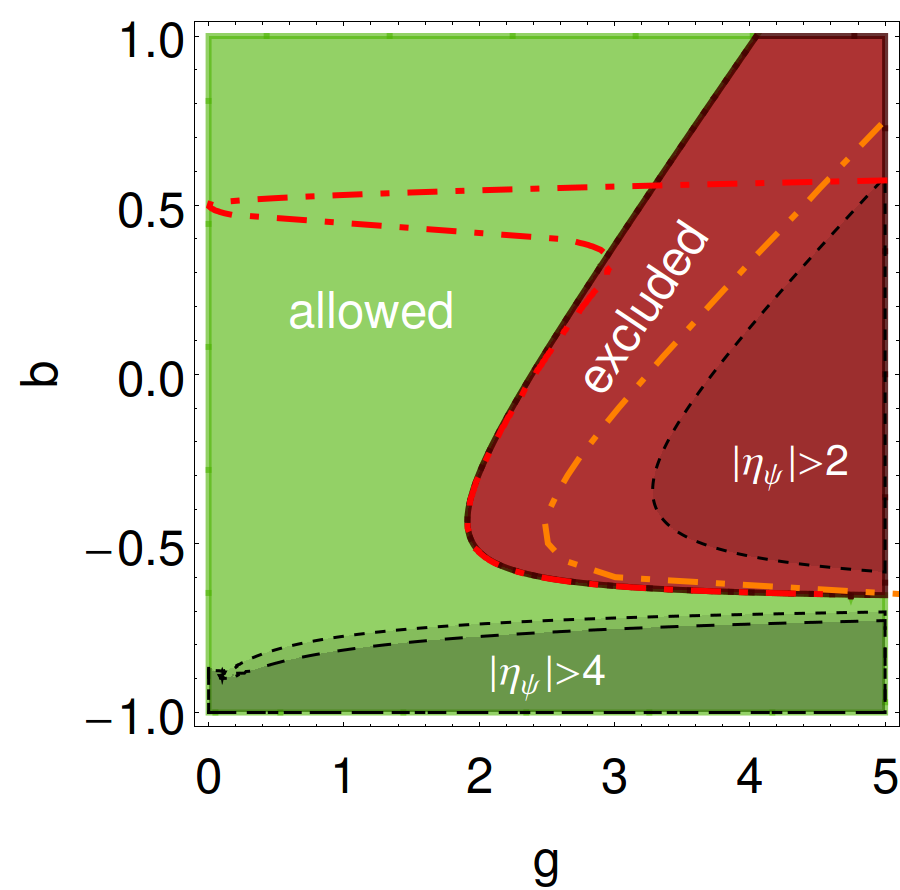}
	\includegraphics[width=0.28\linewidth]{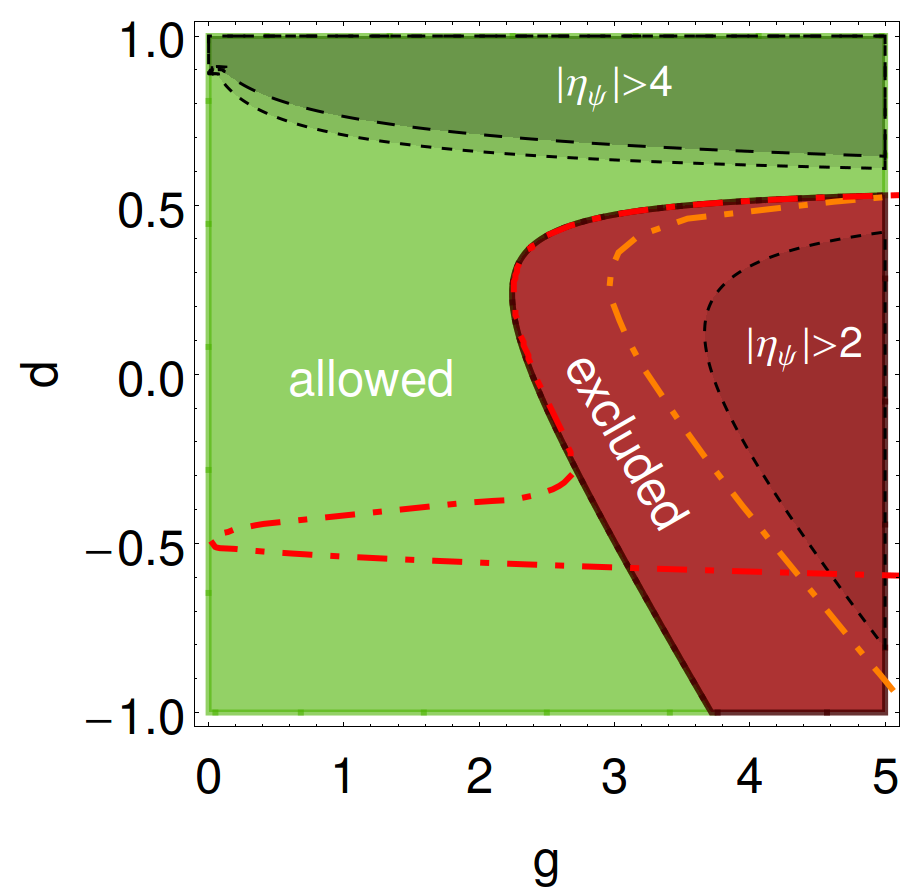}
	\\
	\vspace*{10pt}
	\includegraphics[width=0.28\linewidth]{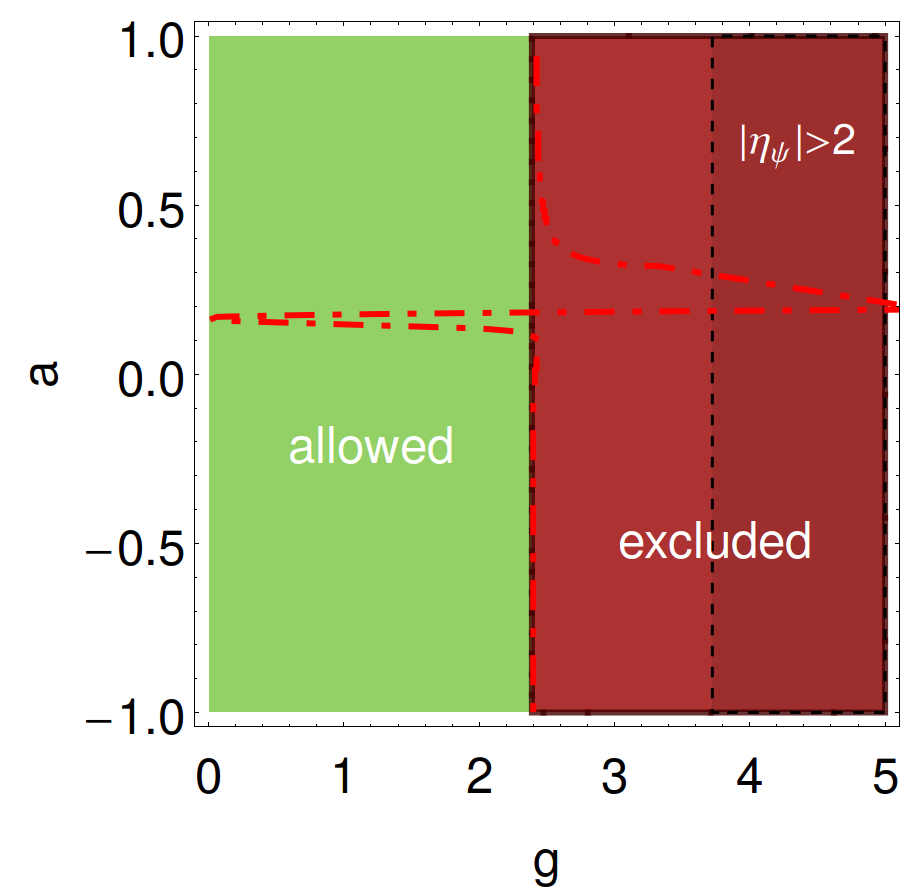}
	\hspace*{10pt}
	\includegraphics[width=0.28\linewidth]{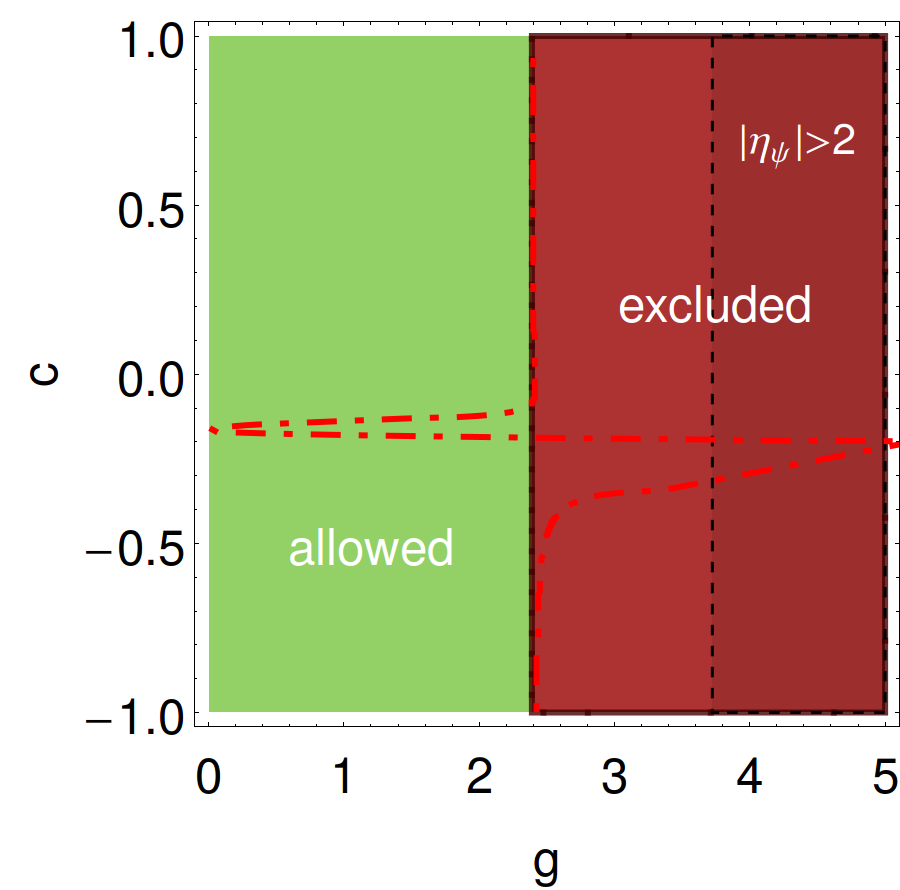}
	\caption{\label{fig:weakGravity_bound_loosendApprox}
	As Fig.~\ref{fig:weakGravity_bound}, including a numerical bound (thick red dot-dashed line) lifting	the following approximations: we include the trace-mode contributions; we solve the full set of anomalous dimensions (including loop-terms); we include effects from $\beta_{\lambda_A}$. For each panel all gravitational couplings not plotted are set to zero. Besides a novel pole that is introduced due to an (unphysical) regime in which the trace-mode dominates the bound is only slightly modified.
	}
\end{figure*}

Building on the simplified analysis that we have presented in the main text, we can now launch a fully-fledged discussion of the flow without resorting to approximations for the gravity modes, and provide a detailed discussion of all results beyond the TT approximation. As a main result of this section, we confirm that the TT approximation employed in the main text is quantitatively adequate except for particular parts of the gravitational coupling space, i.e., for large $\mu_h$ or large higher-order curvature couplings.
In our choice of gauge, $\beta=\alpha=0$, additional contributions to the flow of matter couplings arise from the trace mode only. Including those contributions in Eq.~\eqref{eq:thresholds_start}-\eqref{eq:thresholds_end} and evaluating the integrals for spectrally adjusted cutoffs, introduces a second pole in gravitational parameter space. Besides the spin-2-mode pole
\be
 {\rm pole}_{\rm TT} = \frac{g}{\left(1+\mu_h+b-d\right)}
\ee
now also the trace mode develops a pole at
\be
\label{eq:TTmode_pole}
 {\rm pole}_{\rm Tr} = \frac{g}{\left(3+2\mu_h- 18 (a-c) - 6 (b-d) \right)}\;.
\ee
Apart from dominant behavior of the trace mode, accompanied by possible changes of the overall sign of contributions close to these poles, the results deviate only very slightly from the approximations employed in the main text (cf.~FIG.~\ref{fig:weakGravity_bound_loosendApprox}~\&~\ref{fig:condition_eff}). Since the vicinity of the pole of the trace-mode propagator corresponds to regimes in gravitational parameter space in which the trace-mode is enhanced and dominates over the TT-mode we regard these as unphysical. Physics should be determined by the gauge-independent spin-2-mode and not by the gauge-dependent trace-mode. Setting all other gravitational parameters to zero the trace-mode pole lies at
\begin{align}
	&\mu_{h,\text{pole}}|_{a=b=c=d=0} = -\frac{3}{2}\;, \\
	&a_\text{pole}|_{b=c=d=\mu_h=0} = -c_\text{pole}|_{a=b=d=\mu_h=0} = \frac{1}{6}\;, \\
	&b_\text{pole}|_{a=c=d=\mu_h=0} = -d_\text{pole}|_{a=b=c=\mu_h=0} = \frac{1}{2} \;.
\end{align}
Comparing with FIG.~\ref{fig:weakGravity_bound_loosendApprox} shows that apart from strong pole modifications at these values the results are in good agreement with the spin-2-approximation. We conclude that results obtained in the spin-2-approximation might be sufficient to determine the physical effect of gravity on the matter sector. In particular, gravitational parameters such as those in a Ricci-scalar expansion, i.e., the parameters $a$ and $c$ in our truncation do not -- apart from unphysical pole behavior -- have a strong effect on the matter sector. This justifies the TT-mode approximation used in this paper.

\end{document}